\documentclass[12pt]{article}
\usepackage{eepic}
\usepackage[usenames]{color}
\usepackage{amssymb}

 \def\be{\begin{equation}}
 \def\ee{\end{equation}}
 \def\bea{\begin{eqnarray}}
 \def\eea{\end{eqnarray}}
 \def\a{\alpha}
 \def\b{\beta}
 \def\g{\gamma}
 \def\d{\delta}
 \def\o{\omega}
 \def\s{\sigma}
\def\G{\Gamma}
\def\L{\Lambda}
 
\def\ho{\tilde{\omega}}
\def\tio{\hat{\omega}}
\def\tp{\tilde{p}}
 
\def\p{\partial}
\def\A{\mathcal{A}}
\def\B{\mathcal{B}}
\def\F{\mathcal{F}}
\def\H{\mathcal{H}}
\def\K{\mathcal{K}}
\def\N{\mathcal{N}}
\def\O{\mathcal{O}}
\def\W{\mathcal{W}}
\def\2{\frac{1}{2}}
\def\4{\frac{1}{4}}

\def\ap{a_+}
\def\am{a_-}

\def\np#1{{\sl Nucl.~Phys.~\bf B#1}}
\def\pl#1{{\sl Phys.~Lett.~\bf B#1}}
\def\pr#1{{\sl Phys.~Rev.~\bf D#1}}
\def\prl#1{{\sl Phys.~Rev. Lett.~\bf #1}}
\def\cm#1{{\sl Comm.~Math.~Phys.~\bf #1}}
\def\mpl#1{{\sl Mod.~Phys.~Lett.~\bf A#1}}
\def\cpc#1{{\sl Comp.~Phys.~Comm.~\bf #1}}
\def\anp#1{{\sl Ann.~Phys.~(NY) \bf #1}}
\def\rmp#1{{\sl Rev.~Mod.~Phys. \bf #1}}
\def\cqg#1{{\sl Class.~Quant.~Grav.~\bf #1}}

\catcode`\@=11
\def\@citex[#1]#2{%
\if@filesw \immediate \write \@auxout {\string \citation {#2}}\fi
\@tempcntb\m@ne \let\@h@ld\relax \def\@citea{}%
\@cite{%
  \@for \@citeb:=#2\do {%
    \@ifundefined {b@\@citeb}%
      {\@h@ld\@citea\@tempcntb\m@ne{\bf ?}%
      \@warning {Citation `\@citeb ' on page \thepage \space undefined}}%
      {\@tempcnta\@tempcntb \advance\@tempcnta\@ne%
      \@tempcntb\number\csname b@\@citeb \endcsname \relax%
      \ifnum\@tempcnta=\@tempcntb 
        \ifx\@h@ld\relax%
          \edef \@h@ld{\@citea\csname b@\@citeb\endcsname}%
        \else%
          \edef\@h@ld{\ifmmode{-}\else--\fi\csname b@\@citeb\endcsname}%
        \fi%
      \else
        \@h@ld\@citea\csname b@\@citeb \endcsname%
        \let\@h@ld\relax%
      \fi}%
    \def\@citea{,\penalty\@highpenalty\,}%
  }\@h@ld
}{#1}}

\def\@citeb#1#2{{[#1]\if@tempswa , #2\fi}}
\def\@citeu#1#2{{$^{#1}$\if@tempswa , #2\fi }}
\def\@citep#1#2{{#1\if@tempswa , #2\fi}}

\title{Perturbative calculation of quasi-normal modes of AdS Schwarzschild black holes}
\author{{\bf Suphot Musiri}\footnote{suphot@swu.ac.th}\\
\em Department of Physics,
Srinakharinwirot University, \\
\em Bangkok 10110, Thailand. \\ \\
\and {\bf Scott Ness\footnote{ness@utk.edu}\ \ and George Siopsis}\footnote{siopsis@tennessee.edu}\\
\em Department of Physics and Astronomy,
The University of Tennessee, \\
\em Knoxville, TN 37996 - 1200, USA.
}
\date{October 2005}
\begin{document}

\maketitle
\vspace{-5in}\hfill UTHET-05-0701\vspace{5in}

\abstract{We calculate analytically quasi-normal modes of AdS Schwarzschild black holes including first-order corrections.
We consider massive scalar, gravitational and electromagnetic perturbations.
Our results are in good agreement with numerical calculations.
In the case of electromagnetic perturbations, ours is the first calculation
to provide an analytic expression for quasi-normal frequencies, because the
effective potential vanishes at zeroth order.
We show that the first-order correction is logarithmic.
}
\newpage
\section{Introduction}

Recently there has been a lot of research studying quasi-normal modes of black holes in asymptotically AdS space-times \cite{qnm}.  Understanding these modes may give some insight into the AdS/CFT correspondence.

Here we present a fairly comprehensive study of quasi-normal modes of
AdS Schwarzschild black holes with a metric in $d$ dimensions given by
\be\label{line}
ds^2 = -f(r)dt^2 +\frac{ dr^2 }{ f(r) } +r^2d \Omega_{d-2}^2\ \ ,\ \ \ f(r) = \frac{r^2}{R^2}+1-\frac{2\mu}{r^{d-3}} \;.
\ee
and derive analytical expressions including
first-order corrections. The results are in good agreement with results of
numerical analysis.

In the case of massive perturbations, which we discuss in section~\ref{sect1},
we extend the approach of~\cite{Siopsis} to include black holes of arbitrary
size. We perform an expansion in $1/m$, where $m$ is the mass of the perturbation.
The calculation involves a large amount of cancellations between various terms
resulting in a sensible perturbative expansion of asymptotic expressions for
quasi-normal frequencies.
In section~\ref{sect2}, we discuss gravitational perturbations and obtain the first-order corrections to
the analytic expressions derived in~\cite{CNS,NS} (adapting the monodromy argument
proposed in~\cite{MN} and extended to first order in~\cite{SuSi}).
We find good agreement with numerical results~\cite{CKL}.
In section~\ref{sect3}, we extend the discussion to electromagnetic perturbations.
In this case, the zeroth-order effective potential vanishes rendering the analytic
derivation of quasi-normal frequencies impossible.
We show that including the first-order correction leads to an analytic expression
in agreement with numerical results.
Unlike other types of perturbation, the correction in electromagnetic modes is
logarithmic.
We summarize our conclusions in section~\ref{sect4}.

\section{Massive scalar perturbations}\label{sect1}

In this section we calculate quasi-normal frequencies for massive scalar perturbations of {\em finite} black holes in AdS generalizing
a procedure introduced in~\cite{Siopsis}.
We consider explicitly the five-dimensional case in which the wave equation
reduces to a Heun equation.
Generalizing to higher dimensions is straightforward albeit tedious due to the
increase in singular points.

Using the line element~(\ref{line}) in $d=5$, we obtain the horizon radius
\be \frac{r_H^2}{R^2} = -\frac{1}{2} + \sqrt{\frac{1}{4} + \frac{2\mu}{R^2}} \ee
The wave equation for a massive scalar of mass $m$ is
\be\label{eqn4}
  \frac{1}{r^3}\p_r\left(r^3f(r)\p_r\Phi\right)-\frac{1}{f(r)}\p_t^2\Phi+\frac{1}{r^2}\nabla_\Omega^2\Phi = m^2\Phi\;.
\ee
It is convenient to transform to a dimensionless coordinate
\be
  y= s\left( \frac{2r^2}{R^2} + 1 \right) \ \ , \ \ \ \ s = \frac{1}{\frac{2r_H^2}{R^2} + 1}\;,
\ee
in terms of which the factor $f(r)$ (eq.~(\ref{line}) with $d=5$) reads
\be
  f[r(y)] = \frac{y^2-1}{2s (y-s)}\;.
\ee
We see that $s$ is a parameter describing the size of the black hole.  When $s \to 0$, we approach the large black hole limit ($r_H\to\infty$) and expect to arrive at the results of \cite{Siopsis}.

Separating variables,
\be
  \Phi = e^{-i\o t}Y_{\ell \vec m} (\Omega)\Psi(y) \ ,
\ee
we obtain the radial wave equation expressed in terms of $y$,
\be\label{wew}
  (y-s)(y^2-1)\Psi''+\left(3y^2-1-2 sy\right)\Psi'+\left[\frac{\tio^2}{4}\frac{(y-s)^2}{y^2-1}-\frac{\hat{L}^2}{4}-(y-s)\hat{m}^2\right]\Psi =0
\ee
where we introduced the dimensionless parameters
\be
  \hat\o^2 = 2s\o^2R^2 \;,\ \ \ \hat{L}^2 = 2s\ell(\ell+2) \;,\ \ \ \hat{m} = \frac{mR}{2} \;.
\ee
The singularities of the wave equation are given by
\be
  y=\pm 1,s\;,
\ee
where $y=1$ is the horizon, $y=s$ is the black hole singularity and $y=-1$ is
an unphysical singularity.
In order to bring~(\ref{wew}) into a manageable form,
we need to study the behavior of the wavefunction near the singularities.
Two independent solutions of~(\ref{wew}) are obtained by examining the behavior near the horizon ($y\to 1$),
\be\label{psipm}
  \Psi_\pm \sim (y-1)^{\pm i \frac{\tio}{4} \sqrt{1-s}}\;.
\ee
where $\Psi_+,\Psi_-$ represent outgoing and ingoing waves, respectively.
We will choose $\Psi_-$ for quasi-normal modes. 

Near the singularity $y\to -1$ we obtain a different set of independent solutions
\be
  \Psi \sim (y+1)^{\pm  \frac{\tio}{4} \sqrt{1+s}}\;.
\ee
Since this is an unphysical singularity there is no physical choice.
By studying the behavior at large $r$ ($y\to \infty$),
we find another set of independent solutions which determine the scaling behavior and are given by
\be\label{scale}
  \Psi\sim y^{-h_\pm} \ \ ,\ \ h_\pm=1\pm\sqrt{1+\hat{m}^2}.
\ee
For quasi-normal modes we want the solution to vanish for large $r$ ($y\to\infty$), leading us to choose
\be
  \Psi\sim y^{-h_+} \ .
\ee
%
We may write the solution of~(\ref{wew}) in the form
\be\label{eqn22}
  \Psi = (y-1)^{-i \frac{\tio}{4} \sqrt{1-s}}(y+1)^{- \frac{\tio}{4} \sqrt{1+s}}F(y) \ .
\ee
Substituting this expression into the wave equation~(\ref{wew}),
we obtain an equation for $F(y)$,
\bea\label{eq15}
  (y^2-1)F''+\left\{\left(3-(\sqrt{1+s}+i\sqrt{1-s})\frac{\tio}{2}\right) y
+s+(\sqrt{1+s}-i\sqrt{1-s})\frac{\tio}{2}\right\}F'&&\nonumber\\
  +\left\{\frac{\tio}{2}\left[\left(s+i\sqrt{1-s^2}\right)\frac{\tio}{4}-(\sqrt{1+s}+i\sqrt{1-s})\right]-\hat{m}^2\right\}F &&\nonumber\\
+\frac{1}{y-s}\left\{(s^2-1)F'- \frac{\hat L^2}{4}\, F +\left[(1-s)\sqrt{1+s}-i(1+s)\sqrt{1-s}\right]\frac{\tio}{4}F\right\} &&\nonumber\\
= 0 &&\eea
If we are interested in the limit of large frequencies $\tio$,
we may focus on the region of large $y$~\cite{Siopsis}.
In this case, the last term on the left-hand side of~(\ref{eq15}) is negligible
compared with the other terms and the wave equation simplifies to a hypergeometric equation,
\bea\label{wey}
  (y^2-1)F''+\left\{\left(3-(\sqrt{1+s}+i\sqrt{1-s})\frac{\tio}{2}\right) y
+s+(\sqrt{1+s}-i\sqrt{1-s})\frac{\tio}{2}\right\}F'&&\nonumber\\
  +\left\{\frac{\tio}{2}\left[\left(s+i\sqrt{1-s^2}\right)\frac{\tio}{4}-(\sqrt{1+s}+i\sqrt{1-s})\right]-\hat{m}^2\right\}F &&\nonumber\\
= 0 &&\eea
Two linearly independent solutions of ($\ref{wey}$) are
\be\label{soln}
  \F_1 = F(\ap,\am;c;-x)\;,\ \ \ \F_2 = x^{1-c}F(1+\ap-c,1+\am-c;2-c;-x)\ ,\ \ \ \   x=\frac{y-1}{2} \ ,
\ee
where
\bea\label{apamc}
  a_\pm &=& h_\pm-\left(\sqrt{1+s}+i\sqrt{1-s}\right)\frac{\tio}{4}\ ,\\
  c &=& \frac{3}{2}+\2(s-i\sqrt{1-s}\;\tio) \ .
\eea
Using the transformation properties of hypergeometric functions,
we may re-express the solutions (\ref{soln}) in terms of a new set of independent solutions which match the scaling behavior~(\ref{scale}) for large $r$ $(x\rightarrow\infty)$,
\be\label{eqsolinf}
  \mathcal{K}_\pm = (x+1)^{-a_\pm} F(a_\pm, c-a_\mp ; a_\pm -a_\mp +1; 1/(x+1))
\ee
We ought to choose $\K_+$, since it leads to $\Psi\rightarrow 0$ as $x\rightarrow\infty$.
$\K_+$ may be expressed as a linear combination of $\F_1$ and $\F_2$,
\be
  \mathcal{K}_+ = \mathcal{A}_0 \mathcal{F}_1 + \mathcal{B}_0\mathcal{F}_2,
\ee
where
\be\label{eqn21}
  \mathcal{A}_0 = \frac{\Gamma(1-c)\Gamma(1-a_-+a_+)}{\Gamma(1-a_-)\Gamma(1-c+a_+)}\ \ , \ \ \mathcal{B}_0 = \frac{\Gamma(c-1)\Gamma(1+a_+-a_-)}{\Gamma(a_+)\Gamma(c-a_-)}.
\ee
For the correct behavior at the horizon, we demand
\[
  \mathcal{B}_0 = 0,
\]
which leads to two conditions
\be\label{cond1}
  c-\am = 1-n\ \ ,\ \ n=1,2,3,\dots
\ee
or
\be\label{cond2}
  \ap=1-n\ \ ,\ \ n=1,2,3,\dots
\ee
Eq.~(\ref{cond1}) leads to the zeroth-order frequencies,
\be\label{f1}
   \tio_n = -2(\sqrt{1+s}+i\sqrt{1-s})\, \left[n+h_+-\frac{3}{2}+\frac{s}{2}\right]
\ee
Notice that the phase approaches $\pi/4$ in the large black-hole limit ($r_H\to\infty$ or $s\to 0$), as expected~\cite{Siopsis}.

Using (\ref{cond2}), we find a second set of frequencies given by
\be\label{f2}
  \tio_n=2(\sqrt{1+s}-i\sqrt{1-s})(n+h_+-1) \ .
\ee
Both sets of frequencies, (\ref{f1}) and (\ref{f2}), at leading order agree on the imaginary part
and have opposite real parts.
We shall work with~(\ref{f1}) without loss of generality.
Notice also that the two sets of quasi-normal frequencies match the results
of~\cite{Siopsis} in the large black hole limit ($s\to 0$).

To find the first-order correction to the zeroth-order expression for quasi-normal frequencies~(\ref{f1}),
we shall solve the Heun equation~(\ref{eq15}) perturbatively.
To this end, let us bring it to the form
\be
  \left( \H_0+\H_1\right)F =0 ,
\ee
where ({\em cf.}~eq.~(\ref{wey}))
\bea\label{H0}
  \H_0 &=& \p_y^2+
  \frac{1}{y^2-1} \left\{\left(3-(\sqrt{1+s}+i\sqrt{1-s})\frac{\tio}{2}\right) y+(s+(\sqrt{1+s}-i\sqrt{1-s})\frac{\tio}{2}\right\}\p_y\nonumber\\
  &&+\frac{1}{y^2-1} \left\{ \frac{\tio}{2}\left[(s+i\sqrt{1-s^2})\frac{\tio}{4}-(\sqrt{1+s}+i\sqrt{1-s})\right]-\hat{m}^2\right\},
\eea
and the correction (to be treated as a perturbation) is given by
\be
  \H_1 = \frac{1}{(y^2-1)(y-s)}\left[(s^2-1)\p_y+\left((1-s)\sqrt{1+s}-i(1+s)\sqrt{1-s}\right)\frac{\tio}{4}\right].
\ee
We have neglected the angular momentum contribution for simplicity.
We may expand the wave function as
\be
  F=F_0+F_1+\dots
\ee
where $F_0$ obeys the zeroth-order equation (eq.~(\ref{wey}))
\be
  \H_0F_0=0.
\ee
Solving this equation leads to the zeroth-order expressions for quasi-normal
frequencies~(\ref{f1}).
The first-order equation is
\be
  \H_1F_0+\H_0F_1=0
\ee
We may solve for $F_1$ by using variation of parameters,
\be\label{F1}
  F_1=\K_-\int_x^\infty\frac{\K_+\H_1F_0}{\W}-\K_+\int_x^\infty\frac{\K_-\H_1F_0}{\W}
\ee
where $\K_\pm$ are the two linearly independent solutions (\ref{eqsolinf}) of eq.~(\ref{wey}) and $\W$ is their Wronskian given by
\be
  \W=(a_+-a_-)x^{-c}(1+x)^{c-a_+-a_--1}.
\ee
To study the behavior near the horizon ($x\to 0$),
we may analytically continue the parameters in (\ref{F1}) without affecting the singularity.
For $x\sim 0$, we obtain
\be
  F_1\sim \A_1+x^{1-c}\B_1\ \ ,
\ee
where
\be
  \B_1=\b_-\int_0^\infty\frac{\K_+\H_1F_0}{\W}-\b_+\int_0^\infty\frac{\K_-\H_1F_0}{\W},
\ee
and
\be
  \b_\pm=\frac{\Gamma(c-1)\Gamma(1+a_\pm-a_\mp)}{\Gamma(a_\pm)\Gamma(c-a_\mp)}.
\ee
With our choice (\ref{cond1}), we find
\[
  \B_1=\b_-\int_0^\infty\frac{\K_+\H_1F_0}{\W}.
\]
Therefore the quasi-normal frequencies, to first order, are found as solutions of
\be\label{b0b1}
  \B_0+\B_1=0,
\ee
where $\B_0$ is given by (\ref{eqn21}).

We can now find explicit expressions for the first-order correction to quasi-normal frequencies.
Writing to first order
\be\label{f1a}
   \tio_n = -2(\sqrt{1+s}+i\sqrt{1-s})\, \left[n+h_+-\frac{3}{2}+\frac{s}{2}
-\epsilon_n\right]
\ee
we aim at calculating $\epsilon_n$.
Let us start with the case of $n=1$.
Our quantization condition (\ref{cond1}) becomes $c=a_-$.  This truncates the expansion of the hypergeometric solution (\ref{eqsolinf}) to
\be
  F_0=\K_+=(1+x)^{-a_+}.
\ee
After some algebra, we find
\be
  \B_1=\frac{\b_-}{2(a_+-a_-)}\sum_{k=0}^1\a_k\int_0^\infty dx\frac{x^c(1+x)^{-(c+a_+-a_--k)}}{2x+1-s}\ ,
\ee
where the coefficients, $\a_k$ ($k=0,1$), are given by
\be
  \a_0= -a_+(s^2-1) \ \ ,\ \ \a_1= \left[(s^2-1)+is\sqrt{1-s^2}\right]\left[a_+-a_-+1+s\right].
\ee
Using
\be
  \int_0^\infty dx\frac{x^{\lambda}(1+x)^{-\mu}}{1+\d x}=B(\lambda+1,\mu-\lambda)F(1,\lambda+1;\mu+1;1-\d),
\ee
we find
\bea
  \B_1&=&\frac{B(a_--1,a_+-a_-+1)}{1-s}\left(\frac{a_-}{a_+-a_-}\right)\left[-\frac{\a_0}{2a_+}F(1,a_-+1;a_++1;\frac{s+1}{s-1})\right.\nonumber\\
  &&\left.-\frac{\a_1}{2(a_+-a_--1)}F(1,a_-+1,a_+;\frac{s+1}{s-1})\right].
\eea
Expanding in $1/h_+$ (large mass expansion),
we obtain
\be
  \B_1=B(a_--1,a_+-a_-+1)\left[\frac{s}{4} \left( s^2-1 +is\sqrt{1-s^2}
\right)
-\frac{i}{8h_+}\left( 1+ o(s) \right)\right],
\ee
where we made use of the expansion of a hypergeometric function
\be\label{eqhyp} F(1,\alpha;\beta; z)
= \left(1-\frac{\alpha}{\beta}\ z\right)^{-1}+ \left(\frac{1}{\alpha} - \frac{1}{\beta} \right)
\frac{\alpha^2 z^2}{\beta^2} \left(1-\frac{\alpha}{\beta}\ z\right)^{-3} + \dots\ee
which is valid for large $\alpha$ and $\beta$ ($\sim o(h_+)$).
%
We obtain from (\ref{eqn21}) and (\ref{f1a})
\be
  \B_0=\epsilon_1 B(\am-1,\ap-\am+1)+\dots
\ee
By using (\ref{b0b1}) we find the first-order correction for $n=1$,
\be
  \epsilon_1= -\frac{s}{4} \left( s^2-1 +is\sqrt{1-s^2}
\right)
+\frac{i}{8h_+}\left( 1+ o(s) \right)
\ee
For a finite-size black hole ($s\ne 0$), this is a $o(h_+^0)$ correction
to $n=1$ quasi-normal frequencies. The correction is $o(1/h_+)$ for
an infinite-size black hole ($s=0$)~\cite{Siopsis}.
It should be pointed out that the calculation of $\epsilon_1$ involved
cancellation of $o(h_+)$ terms.
For a general $n$, one obtains expressions $o(h_+^n)$. Non-trivial
cancellations occur between various terms involving hypergeometric functions
and after the dust settles, one arrives at the general expression
\be
  \epsilon_n = -\frac{s}{4n} \left( s^2-1+is\sqrt{1-s^2} \right) +
\frac{i (2-1/n)}{8h_+} \left( 1 + o(s) \right) \ \ ,\ \ n=1,2,3,\dots
\ee
which is $o(h_+^0)$ for finite-size black holes and $o(1/h_+)$ for infinite-size
black holes,
\be
  \epsilon_n = \frac{i}{4} \left( 1 - \frac{1}{2n} \right) \frac{1}{h_+} \;.
\ee
We have been unable to provide an analytical proof of the above results for general $n$
but have verified them for several $n$ using {\sf Mathematica}. 

\section{Gravitational perturbations}\label{sect2}

In this section we discuss gravitational perturbations.
For massless perturbations, the method discussed in section~\ref{sect1} is
not directly applicable.
Instead, we extend the procedure of~\cite{CNS,NS} to include first-order
corrections to analytical expressions for quasi-normal frequencies.
Our results are in good agreement with numerical results~\cite{CKL}.

The radial wave equation for gravitational perturbations in the black-hole
background~(\ref{line}) can be cast into a Schr\"odinger-like form,
\be\label{sch}
  -\frac{d^2\Psi}{dr_*^2}+V[r(r_*)]\Psi =\o^2\Psi \;,
\ee
in terms of the tortoise coordinate defined by
\be\label{tortoise}
  \frac{dr_*}{dr} = \frac{1}{f(r)}\;.
\ee
The potential $V$ is determined by the type of perturbation and may be
deduced from the Master Equation derived in~\cite{IK}.
For tensor, vector and scalar perturbations, we obtain, respectively,~\cite{NS}
\be\label{eqVT} V_{\mathsf{T}} (r) = f(r) \left\{ \frac{\ell (\ell +d-3)}{r^2} + \frac{(d-2)(d-4) f(r)}{4r^2} + \frac{(d-2) f'(r)}{2r} \right\} \ee
\be\label{eqVV} V_{\mathsf{V}}(r) = f(r) \left\{ \frac{\ell (\ell +d-3)}{r^2} + \frac{(d-2)(d-4) f(r)}{4r^2} - \frac{r f'''(r)}{2(d-3)} \right\} \ee
\bea\label{eqVS} V_{\mathsf{S}}(r) &=& \frac{f(r)}{4r^2} \left[ \ell (\ell +d-3) - (d-2) + \frac{(d-1)(d-2)\mu}{r^{d-3}} \right]^{-2} \nonumber\\
&\times& \Bigg\{ \frac{d(d-1)^2(d-2)^3 \mu^2}{R^2r^{2d-8}}
- \frac{6(d-1)(d-2)^2(d-4)[\ell (\ell+d-3) - (d-2)] \mu}{R^2r^{d-5}}\nonumber\\
&& + \frac{(d-4)(d-6)[\ell (\ell+d-3) - (d-2)]^2 r^2}{R^2} +
\frac{2(d-1)^2(d-2)^4 \mu^3}{r^{3d-9}}\nonumber\\
&& + \frac{4(d-1)(d-2)(2d^2-11d+18)[\ell (\ell+d-3) - (d-2)]\mu^2}{r^{2d-6}}\nonumber\\
&& + \frac{(d-1)^2(d-2)^2(d-4)(d-6)\mu^2}{r^{2d-6}}
- \frac{6(d-2)(d-6)[\ell (\ell+d-3) - (d-2)]^2 \mu}{r^{d-3}}\nonumber\\
&& - \frac{6(d-1)(d-2)^2(d-4)[\ell (\ell+d-3) - (d-2)] \mu}{r^{d-3}}\nonumber\\
&& + 4 [\ell (\ell+d-3) - (d-2)]^3 + d(d-2) [\ell (\ell+d-3) - (d-2)]^2 \Bigg\} \eea
Evidently, the potential always vanishes at the horizon ($V(r_H) = 0$, since $f(r_H)=0$) regardless of the type of perturbation.

Near the black hole singularity ($r\sim 0$), the tortoise coordinate~(\ref{tortoise}) may be expanded as
\be\label{x0}
r_* = -\frac{1}{(d-2)}\frac{r^{d-2}}{2\mu} - \frac{1}{(2d-5)}\frac{r^{2d-5}}{(2\mu)^2} +\dots
\ee
where we have kept the second term in the expansion of $r$ and have chosen the
integration constant so that $r_*=0$ at $r=0$.
Using~(\ref{x0}), we may expand the potential near the black hole singularity in the three different cases
(eqs.~(\ref{eqVT}), (\ref{eqVV}) and (\ref{eqVS})), respectively as
\be\label{eqVT0} V_{\mathsf{T}} = -\frac{1}{4r_*^2}+\frac{\A_{\mathsf{T}} }{[-2(d-2)\mu]^{\frac{1}{d-2}}} r_*^{-\frac{d-1}{d-2}} + \dots \, , \ \ \ \
  \A_{\mathsf{T}} = \frac{(d-3)^2}{2(2d-5)}+\frac{\ell(\ell+d-3)}{d-2},
\ee
\be\label{eqVV0} V_{\mathsf{V}} = \frac{3}{4r_*^2}+\frac{\A_{\mathsf{V}} }{[-2(d-2)\mu]^{\frac{1}{d-2}}} r_*^{-\frac{d-1}{d-2}} + \dots \ \ , \ \ \ \
\A_{\mathsf{V}} = \frac{d^2-8d+13}{2(2d-15)} + \frac{\ell (\ell +d-3)}{d-2}\ee
and
\be\label{eqVS0}
  V_{\mathsf{S}} =  -\frac{1}{4r_*^2}+\frac{\A_{\mathsf{S}} }{[-2(d-2)\mu]^{\frac{1}{d-2}}} r_*^{-\frac{d-1}{d-2}} + \dots \, ,
\ee
where
\be\label{eqVS0a}
  \A_{\mathsf{S}} = \frac{ (2d^3-24d^2+94d-116)}{4(2d-5)(d-2)}+\frac{ (d^2-7d+14)[ \ell(\ell+d-3)-(d-2)]}{(d-1)(d-2)^2}
\ee
We have included only the terms which contribute to the order we are interested in.
We may summarize the behavior of the potential near the origin by
\be\label{eqV0} V= \frac{j^2 -1}{4r_*^2}+\A\, r_*^{-\frac{d-1}{d-2}} + \dots \ee
where $j=0$ ($2$) for scalar and tensor (vector) perturbations
and the constant coefficient $\mathcal{A}$ can be found from eqs.(\ref{eqVT0}),
(\ref{eqVV0}), (\ref{eqVS0}) and (\ref{eqVS0a}) in the various cases.
Throughout the calculation, we shall pretend that $j$ is not an integer.
At the end of the calculation, we shall let $j\to 0,2$, as appropriate.

After rescaling the tortoise coordinate $(z=\o r_*)$, the Schr\"odinger-like wave equation~(\ref{sch}) with the potential~(\ref{eqV0}) becomes
\be\label{sch-eqn}
  -\frac{d^2\Psi}{dz^2}+\left[\frac{j^2-1}{4z^2}-1\right]\Psi =-\A
\; \o^{-\frac{d-3}{d-2}} \, z^{-\frac{d-1}{d-2}}\Psi \, ,
\ee
In the large frequency limit, we may treat the right-hand side of (\ref{sch-eqn}) as a correction.  This will allow us to to solve the equation perturbatively.  We may re-express (\ref{sch-eqn}) as
\be\label{we-h}
  \left( \H_0+\o^{-\frac{d-3}{d-2}} \, \H_1 \right) \Psi =0,
\ee
where
\be\label{H0-H1}
  \H_0= \frac{d^2}{dz^2}-\left[\frac{j^2-1}{4z^2}-1\right]\ \ ,\ \ \H_1=-\A
\; z^{-\frac{d-1}{d-2}}.
\ee
By treating $\H_1$ as a perturbation, we may expand the wave function
\be\label{expandwf}
  \Psi(z)=\Psi_0(z)+\o^{-\frac{d-3}{d-2}} \, \Psi_1(z)+\dots
\ee
and solve (\ref{we-h}) perturbatively.
The zeroth-order wave equation,
\be\label{we-0}
  \H_0\Psi_0(z)=0,
\ee
may be solved in terms of Bessel functions,
\be\label{soln_0}
  \Psi_0(z)=A_1\sqrt{z}\, J_{\frac{j}{2}}(z)+A_2 \sqrt{z}\, N_{\frac{j}{2}}(z).
\ee
For large $z$, it behaves as
\bea \label{soln-0-0}
  \Psi_0(z)&\sim&  \sqrt{\frac{2}{\pi}}\left[A_1\cos(z-\a_+)+A_2\sin(z-\a_+)\right],\nonumber\\
  &=&\frac{1}{\sqrt{2\pi}}(A_1-iA_2)e^{-i\a_+}e^{iz} + \frac{1}{\sqrt{2\pi}}(A_1+iA_2)e^{+i\a_+}e^{-iz}.
\eea
where $\a_\pm = \frac{\pi}{4}(1\pm j)$.

Next, we study the behavior of the wavefunction at large $r$.
In this region,
the tortoise coordinate~(\ref{tortoise}) may be expanded as
\be \label{x_infty}
r_* -\bar r_* = -\frac{R^2}{r}+\frac{1}{3}\frac{R^4}{r^3} +\dots
\ee
The integration constant is readily deduced
from the definition~(\ref{tortoise}) of the tortoise coordinate,
\be\label{x_infty0} \bar r_* = \int_0^\infty \frac{dr}{f(r)} \ee
The potential (eqs.~(\ref{eqVT}), (\ref{eqVV}) and (\ref{eqVS})) for large $r$ may be expanded as
\be V = \frac{j_\infty^2-1}{4(r_*-\bar r_*)^2} + \dots \ee
where $j_\infty = d-1$, $d-3$ and $d-5$ for tensor, vector and scalar perturbations,
respectively.
The Schr\"odinger-like wave equation~(\ref{sch}) in the region of large $r$ becomes
\be\label{2-2}
  -\frac{d^2\Psi}{dr_*^2}+\left[\frac{j_\infty^2-1}{4(r_*-\bar r_*)^2}-\o^2\right]\Psi=0
\ee
Since the potential does not vanish as $r\to\infty$,
the wavefunction ought to vanish there.
Imposing this boundary condition yields the acceptable solution to eq.~(\ref{2-2}),
\be\label{soln-0-infty}
  \Psi(r_*) = B\sqrt{\o(r_*-\bar r_*)}\; J_{\frac{j_\infty}{2}}(\o (r_*-\bar r_*)) \ .
\ee
Notice that $\Psi \to 0$ as $r_*\to \bar r_*$, as desired.
Asymptotically, it behaves as
\be\label{eq54} \Psi(r_*) \sim \sqrt{\frac{2}{\pi}}\, B\cos\left[ \o(r_*-\bar r_*)+\b\right] \ , \ \ \ \
\b =\frac{\pi}{4}(1+ j_\infty) \ee
By matching this expression to the asymptotic behavior (\ref{soln-0-0}) of the
solution in the vicinity of the black-hole singularity along the Stokes line $\Im z = \Im (\o r_*) = 0$, we find a constraint on the coefficients $A_1,\ A_2$,
\be \label{constraint_1}
  A_1\tan(\o \bar r_* -\b -\a_+)-A_2=0.
\ee
A second constraint is obtained by imposing the boundary condition
\be  \Psi(z) \sim e^{iz}\ \ , \ \ \ \ z\to -\infty\label{bc-0}\ ,
\ee
at the horizon.
To this end, we need to analytically continue the wavefunction near the
origin to negative values of $z$.
A rotation of $z$ by $-\pi$ corresponds to a rotation by $-\frac{\pi}{d-2}$ near the origin in the complex $r$-plane, on account of~(\ref{x0}).
Since near the origin, $J_\nu (z) \sim z^\nu$ (multiplied by an even holographic function of $z$) and using the identity
\be\label{eqJN}
  N_\nu(z)=\cot \pi\nu\, J_\nu(z)-\csc \pi\nu\, J_{-\nu}(z) \ ,
\ee
we deduce
\be\label{eqBrot} J_\nu (e^{-i\pi} z) = e^{-i\pi\nu} J_\nu (z) \ , \ \ \ \
N_\nu (e^{-i\pi} z) = e^{i\pi\nu} N_\nu - 2i\cos \pi\nu\, J_\nu (z)\ee
Thus for $z<0$, the wavefunction~(\ref{soln_0}) changes to
\be\label{soln_0r}
  \Psi_0(z)= e^{-i\pi(j+1)/2} \sqrt{-z}\, \left\{ \left[ A_1 -i (1+e^{i\pi j}) A_2 \right]\, J_{\frac{j}{2}}(-z)+A_2 e^{i\pi j} \, N_{\frac{j}{2}}(-z) \right\} \ .
\ee
whose asymptotic behavior is given by
\be \Psi \sim \frac{e^{-i\pi(j+1)/2}}{\sqrt{2\pi}} \left[ A_1-i(1+2e^{j\pi i}) A_2\right]\, e^{-iz}+\frac{e^{-i\pi(j+1)/2}}{\sqrt{2\pi}} \left[ A_1-iA_2\right]\, e^{iz} \ee
Imposing the boundary condition~(\ref{bc-0}) at the horizon, we deduce the constraint
\be\label{constraint_2}
  A_1 -i(1+2e^{j\pi i}) A_2 = 0.
\ee
The two constraints (\ref{constraint_1}) and (\ref{constraint_2}) are compatible provided
\be\label{eqcomp}
  \left| \begin{array}{cc}  1 &  -i(1+2e^{j\pi i}) \\
                          \tan(\o \bar r_*-\b-\a_+) & -1 \end{array}   \right| = 0,
\ee
which yields
the quasi-normal frequencies~\cite{NS}
\be\label{omega-0}
  \o \bar r_* =\frac{\pi}{4} (2+j+ j_\infty)-\tan^{-1} \frac{i}{1+2e^{j\pi i}} +n\pi \ .
\ee
These are zeroth-order expressions deduced from the zeroth-order wave equation~(\ref{we-0}).

%
Next, we calculate the first-order correction to the asymptotic expressions~(\ref{omega-0}) for quasi-normal frequencies.
We begin by focusing on the region near the black-hole singularity ($r\sim 0$).
To first order, the wave equation (\ref{we-h}) becomes 
\be\label{1stwe1}
  \H_0\Psi_1+\H_1\Psi_0=0 \ ,
\ee
where $\H_0$ and $\H_1$ are given in eq.~(\ref{H0-H1}).
The solution is
\be\label{soln_1}
  \Psi_1(z) = \sqrt{z}\, N_{\frac{j}{2}}(z)\int_0^z dz'\frac{\sqrt{z'}\, J_{\frac{j}{2}}(z')
\mathcal{H}_1\Psi_0(z') }{ \W } -  \sqrt{z}\, J_{\frac{j}{2}}(z)\int_0^z dz'\frac{\sqrt{z'}\, N_{\frac{j}{2}}(z') \mathcal{H}_1\Psi_0(z') }{ \W },
\ee
written in terms of the two linearly independent solutions~(\ref{soln_0}) of the zeroth-order eq.~(\ref{we-0}). $\W = 2/\pi$ is their Wronskian.
Using~(\ref{soln_0}) and (\ref{soln_1}), we may express the solution to the wave equation (\ref{we-h}) up to first order (eq.~(\ref{expandwf})) explicitly as
\be\label{soln1st0}
  \Psi(z)=\left\{A_1[1-b(z)] -A_2a_2(z)\right\}\sqrt{z} J_{\frac{j}{2}}(z) +\left\{A_2[1+b(z)]+A_1a_1(z)\right\}\sqrt{z} N_{\frac{j}{2}}(z)
\ee
where the functions $a_1(z)$, $a_2(z)$ and $b(z)$ are given by
\bea
  a_1(z) &=& \frac{\pi\A}{2} \, \omega^{-\frac{d-3}{d-2}}\, \int_0^z dz'\;{z'}^{-\frac{1}{d-2}}J_{\frac{j}{2}}(z') J_{\frac{j}{2}}(z'),\\
  a_2(z) &=& \frac{\pi\A}{2} \, \omega^{-\frac{d-3}{d-2}}\,  \int_0^z dz'\;{z'}^{-\frac{1}{d-2}}N_{\frac{j}{2}}(z') N_{\frac{j}{2}}(z'),\\
  b(z) &=& \frac{\pi\A}{2} \, \omega^{-\frac{d-3}{d-2}}\,  \int_0^z dz'\;{z'}^{-\frac{1}{d-2}}J_{\frac{j}{2}}(z') N_{\frac{j}{2}}(z') \ ,
\eea
respectively. The coefficient $\A$ is defined in eq.~(\ref{H0-H1}) and depends on the type of perturbation.
The wavefunction (\ref{soln1st0}) behaves asymptotically as
\be\label{1stsoln}
  \Psi(z)\sim \sqrt{\frac{2}{\pi}}\, [A_1' \cos(z-\a_+)+ A_2' \sin(z-\a_+)]\ ,
\ee
where
\be\label{eq91} A_1' = [1-\bar b]A_1-\bar a_2 A_2\ \ , \ \ \ \
A_2' = [1+\bar b]A_2+\bar a_1 A_1\ee
and we introduced the notation
\be \bar a_1 = a_1(\infty)\ \ , \ \ \ \ \bar a_2 = a_2(\infty)\ \ , \ \ \ \ \bar b = b(\infty) \ . \ee
By matching this to the asymptotic expression~(\ref{eq54}),
we obtain
\be\label{newconstr1} A_1' \tan (\o \bar r_* -\beta -\alpha_+) - A_2' = 0\ee
correcting the zeroth-order constraint~(\ref{constraint_1}).
Using~(\ref{eq91}), the first-order constraint~(\ref{newconstr1}) in terms of $A_1$ and $A_2$ reads
\be\label{newconstr1a} [ (1-\bar b)\tan (\o \bar r_* -\beta -\alpha_+)- \bar a_1 ]A_1 -[1+\bar b +\bar a_2 \tan (\o \bar r_* -\beta -\alpha_+)]A_2 = 0\ee
To find the first-order correction to the second constraint~(\ref{constraint_2}),
we need to approach the horizon. This entails a rotation by $-\pi$ in the
$z$-plane.
From the small-$z$ behavior of a Bessel function, $J_\nu (z) \sim z^\nu$, and
using the identity~(\ref{eqJN}),
we deduce after some algebra
\bea a_1 (e^{-i\pi} z) &=& e^{-i\pi \frac{d-3}{d-2}} e^{-i\pi j} a_1 (z)\ , \nonumber\\
a_2 (e^{-i\pi} z) &=& e^{-i\pi \frac{d-3}{d-2}} \left[ e^{i\pi j} a_2(z)
- 4 \cos^2 \frac{\pi j}{2} a_1(z) - 2i (1+e^{i\pi j} ) b (z) \right]\ , \nonumber\\
b (e^{-i\pi} z) &=& e^{-i\pi \frac{d-3}{d-2}} \left[ b(z) -i (1+e^{-i\pi j}) a_1(z) \right]\eea
From these expressions and eq.~(\ref{eqBrot}), we arrive at a modified expression for the wavefunction~(\ref{soln1st0}) valid for $z<0$.
In the limit $z\to -\infty$, we obtain
\be\label{soln1st0r}
\Psi(z) \sim -i e^{-ij\pi/2} B_1 \cos(-z-\alpha_+)
-i e^{ij\pi/2} B_2\sin(-z-\alpha_+)
\ee
where
\bea
B_1 &=& 
   A_1 -A_1e^{-i\pi\frac{d-3}{d-2}}[{\bar b}-i(1+e^{-i\pi j}){\bar a}_1]
\nonumber \\
& & -A_2e^{-i\pi\frac{d-3}{d-2}} \left[ e^{+i\pi j}{\bar a}_2-4\cos^2\frac{\pi j}{2}{\bar a}_1-2i(1+e^{+i\pi j}){\bar b} \right] \nonumber\\
  & & -i (1+e^{i\pi j})\left[ A_2 +A_2e^{-i\pi\frac{d-3}{d-2}}[{\bar b}-i(1+e^{-i\pi j}){\bar a}_1]+A_1 e^{-i\pi\frac{d-3}{d-2}} e^{-i\pi j}{\bar a}_1 \right] \nonumber\\
B_2 &=& A_2+A_2e^{-i\pi\frac{d-3}{d-2}}[{\bar b}-i(1+e^{-i\pi j}){\bar a}_1]+A_1e^{-i\pi\frac{d-3}{d-2}}e^{-i\pi j}{\bar a}_1
\eea
By imposing the boundary condition~(\ref{bc-0}) at the horizon, we obtain
\be\label{newconstr2} [1-e^{-i\pi\frac{d-3}{d-2}}(i
\bar a_1 +\bar b)]A_1 -[i(1+2e^{i\pi j})+e^{-i\pi\frac{d-3}{d-2}}((1+e^{i\pi j})\bar a_1 +e^{i\pi j} \bar a_2-i\bar b)
]A_2 = 0 \ee
correcting the zeroth-order constraint~(\ref{constraint_2}).
For compatibility of the two first-order constraints, (\ref{newconstr1a}) and
(\ref{newconstr2}), we need
\be\label{eq99}
  \left| \begin{array}{cc}  1+\bar b+\bar a_2\tan (\o \bar r_* -\beta -\alpha_+) & i(1+2e^{i\pi j})+e^{-i\pi\frac{d-3}{d-2}}((1+e^{i\pi j})\bar a_1 +e^{i\pi j} \bar a_2-i\bar b) \\
                          (1-\bar b)\tan (\o \bar r_* -\beta -\alpha_+)- \bar a_1 & 1-e^{-i\pi\frac{d-3}{d-2}}(i
\bar a_1 +\bar b) \end{array}   \right| = 0
\ee
Solving~(\ref{eq99}), we arrive at the first-order expression for quasi-normal frequencies,
\bea\label{eqo1st}
\omega {\bar r}_* &=& \frac{\pi}{4}(2+j+j_{\infty}) +\frac{1}{2i}\ln 2+n\pi \nonumber\\
   & & -\frac{1}{8}\left\{ 6i\bar b -2i e^{-i\pi\frac{d-3}{d-2}} \bar b  -9\bar a_1+e^{-i\pi\frac{d-3}{d-2}}{\bar a}_1 +{\bar a}_2 - e^{-i\pi\frac{d-3}{d-2}}{\bar a}_2 \right\}
\eea
where we took the limit of interest $j\to 0,2$ wherever it was unambiguous, in
order to simplify the notation.
Using
\be\label{JJ}
  \int_0^\infty dx\;x^{-\lambda}J_\mu(x)J_\nu(x)=\frac{\G(\lambda)\G(\frac{\nu +\mu+1-\lambda}{2})}{2^\lambda \G(\frac{-\nu +\mu+1+\lambda}{2})\G(\frac{\nu-\mu+1+\lambda}{2})\Gamma(\frac{\nu+\mu+1+\lambda}{2})}\ ,
\ee
we obtain explicit expressions for the first-order coefficients,
\bea\label{eqab} \bar a_1 &=& \frac{\pi\mathcal{A}}{4} \left(\frac{n\pi}{2\bar r_*}\right)^{-\frac{d-3}{d-2}} \frac{\G(\frac{1}{d-2})\G(\frac{j}{2}+\frac{d-3}{2(d-2)})}{\G^2(\frac{d-1}{2(d-2)})\Gamma(\frac{j}{2}+\frac{d-1}{2(d-2)})}\nonumber \\
\bar a_2 &=& \left[ 1+2\cot \frac{\pi (d-3)}{2(d-2)} \cot \frac{\pi}{2} \left( -j+\frac{d-3}{d-2}\right) \right]\bar a_1 \nonumber \\
\bar b &=& -\cot \frac{\pi (d-3)}{2(d-2)}\ \bar a_1 \eea
where we used the identity $\Gamma (x) \Gamma(1-x) = \frac{\pi}{\sin \pi x}$.
We also set $\o = n\pi /\bar r_*$, since corrections contribute to higher than first order.
Notice that these expressions are well-defined when $j$ becomes an integer.
Thus, the first-order correction is $\sim o(n^{-\frac{d-3}{d-2}})$.

Next, we compare with numerical results in four dimensions~\cite{CKL}.
It is convenient to set the AdS radius $R=1$. From~(\ref{line}), the
radius of the horizon $r_H$ is related to the black hole parameter $\mu$ by
\be\label{eqmur} 2\mu = r_H^3 + r_H \ee
for $d=4$.
$f(r)$ has two more (complex) roots, $r_-$ and its complex conjugate, where
\be\label{eqrm} r_- = e^{i\pi/3} \left( \sqrt{\mu^2 + \frac{1}{27}} - \mu \right)^{1/3}
- e^{-i\pi/3} \left( \sqrt{\mu^2 + \frac{1}{27}} + \mu \right)^{1/3}
\ee
The integration constant in the tortoise coordinate~(\ref{x_infty0}) is
\be\label{eqrs} \bar r_* = \int_0^\infty \frac{dr}{f(r)} = - \frac{r_-}{3r_-^2+1} \ln \frac{r_-}{r_H}
- \frac{r_-^*}{3r_-^{*2}+1} \ln \frac{r_-^*}{r_H}
\ee
Despite appearances, this is not a real number, because we ought to define
arguments as $0\le \arg r < 2\pi$.

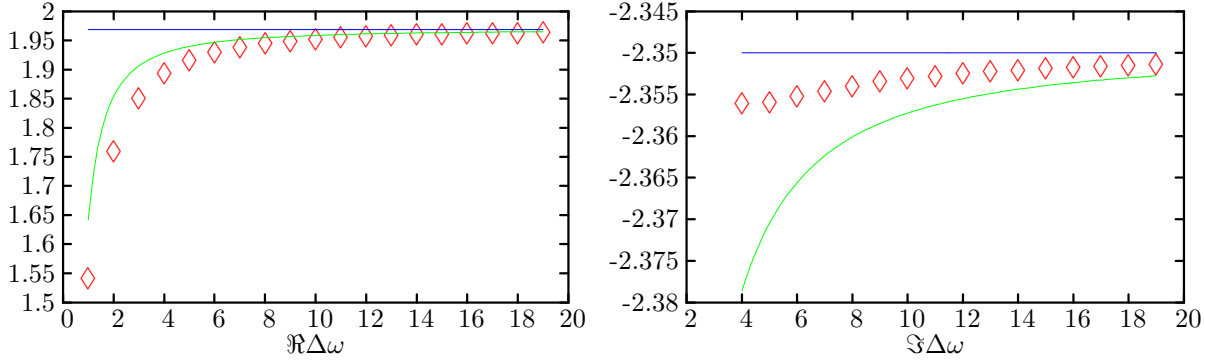
\begin{figure}
\begin{center}
\setlength{\unitlength}{0.0750pt}
\begin{picture}(3000,1800)(0,0)
\footnotesize
\color{black}
\thicklines \path(328,249)(369,249)
\thicklines \path(2876,249)(2835,249)
\put(287,249){\makebox(0,0)[r]{ 1.5}}
\thicklines \path(328,396)(369,396)
\thicklines \path(2876,396)(2835,396)
\put(287,396){\makebox(0,0)[r]{ 1.55}}
\thicklines \path(328,543)(369,543)
\thicklines \path(2876,543)(2835,543)
\put(287,543){\makebox(0,0)[r]{ 1.6}}
\thicklines \path(328,690)(369,690)
\thicklines \path(2876,690)(2835,690)
\put(287,690){\makebox(0,0)[r]{ 1.65}}
\thicklines \path(328,837)(369,837)
\thicklines \path(2876,837)(2835,837)
\put(287,837){\makebox(0,0)[r]{ 1.7}}
\thicklines \path(328,984)(369,984)
\thicklines \path(2876,984)(2835,984)
\put(287,984){\makebox(0,0)[r]{ 1.75}}
\thicklines \path(328,1130)(369,1130)
\thicklines \path(2876,1130)(2835,1130)
\put(287,1130){\makebox(0,0)[r]{ 1.8}}
\thicklines \path(328,1277)(369,1277)
\thicklines \path(2876,1277)(2835,1277)
\put(287,1277){\makebox(0,0)[r]{ 1.85}}
\thicklines \path(328,1424)(369,1424)
\thicklines \path(2876,1424)(2835,1424)
\put(287,1424){\makebox(0,0)[r]{ 1.9}}
\thicklines \path(328,1571)(369,1571)
\thicklines \path(2876,1571)(2835,1571)
\put(287,1571){\makebox(0,0)[r]{ 1.95}}
\thicklines \path(328,1718)(369,1718)
\thicklines \path(2876,1718)(2835,1718)
\put(287,1718){\makebox(0,0)[r]{ 2}}
\thicklines \path(328,249)(328,290)
\thicklines \path(328,1718)(328,1677)
\put(328,166){\makebox(0,0){ 0}}
\thicklines \path(583,249)(583,290)
\thicklines \path(583,1718)(583,1677)
\put(583,166){\makebox(0,0){ 2}}
\thicklines \path(838,249)(838,290)
\thicklines \path(838,1718)(838,1677)
\put(838,166){\makebox(0,0){ 4}}
\thicklines \path(1092,249)(1092,290)
\thicklines \path(1092,1718)(1092,1677)
\put(1092,166){\makebox(0,0){ 6}}
\thicklines \path(1347,249)(1347,290)
\thicklines \path(1347,1718)(1347,1677)
\put(1347,166){\makebox(0,0){ 8}}
\thicklines \path(1602,249)(1602,290)
\thicklines \path(1602,1718)(1602,1677)
\put(1602,166){\makebox(0,0){ 10}}
\thicklines \path(1857,249)(1857,290)
\thicklines \path(1857,1718)(1857,1677)
\put(1857,166){\makebox(0,0){ 12}}
\thicklines \path(2112,249)(2112,290)
\thicklines \path(2112,1718)(2112,1677)
\put(2112,166){\makebox(0,0){ 14}}
\thicklines \path(2366,249)(2366,290)
\thicklines \path(2366,1718)(2366,1677)
\put(2366,166){\makebox(0,0){ 16}}
\thicklines \path(2621,249)(2621,290)
\thicklines \path(2621,1718)(2621,1677)
\put(2621,166){\makebox(0,0){ 18}}
\thicklines \path(2876,249)(2876,290)
\thicklines \path(2876,1718)(2876,1677)
\put(2876,166){\makebox(0,0){ 20}}
\color{black}
\thicklines \path(328,249)(2876,249)(2876,1718)(328,1718)(328,249)
\put(1602,42){\makebox(0,0){$\Re\Delta\omega$}}
\color{red}
\put(455,371){\makebox(0,0){$\Diamond$}}
\put(583,1010){\makebox(0,0){$\Diamond$}}
\put(710,1278){\makebox(0,0){$\Diamond$}}
\put(838,1403){\makebox(0,0){$\Diamond$}}
\put(965,1470){\makebox(0,0){$\Diamond$}}
\put(1092,1510){\makebox(0,0){$\Diamond$}}
\put(1220,1537){\makebox(0,0){$\Diamond$}}
\put(1347,1555){\makebox(0,0){$\Diamond$}}
\put(1475,1567){\makebox(0,0){$\Diamond$}}
\put(1602,1577){\makebox(0,0){$\Diamond$}}
\put(1729,1584){\makebox(0,0){$\Diamond$}}
\put(1857,1590){\makebox(0,0){$\Diamond$}}
\put(1984,1595){\makebox(0,0){$\Diamond$}}
\put(2112,1599){\makebox(0,0){$\Diamond$}}
\put(2239,1602){\makebox(0,0){$\Diamond$}}
\put(2366,1604){\makebox(0,0){$\Diamond$}}
\put(2494,1607){\makebox(0,0){$\Diamond$}}
\put(2621,1608){\makebox(0,0){$\Diamond$}}
\put(2749,1610){\makebox(0,0){$\Diamond$}}
\color{blue}
\thinlines \path(455,1627)(455,1627)(479,1627)(502,1627)(525,1627)(548,1627)(571,1627)(594,1627)(618,1627)(641,1627)(664,1627)(687,1627)(710,1627)(733,1627)(757,1627)(780,1627)(803,1627)(826,1627)(849,1627)(872,1627)(896,1627)(919,1627)(942,1627)(965,1627)(988,1627)(1011,1627)(1034,1627)(1058,1627)(1081,1627)(1104,1627)(1127,1627)(1150,1627)(1173,1627)(1197,1627)(1220,1627)(1243,1627)(1266,1627)(1289,1627)(1312,1627)(1336,1627)(1359,1627)(1382,1627)(1405,1627)(1428,1627)(1451,1627)(1475,1627)(1498,1627)(1521,1627)(1544,1627)(1567,1627)(1590,1627)
\thinlines \path(1590,1627)(1614,1627)(1637,1627)(1660,1627)(1683,1627)(1706,1627)(1729,1627)(1753,1627)(1776,1627)(1799,1627)(1822,1627)(1845,1627)(1868,1627)(1892,1627)(1915,1627)(1938,1627)(1961,1627)(1984,1627)(2007,1627)(2031,1627)(2054,1627)(2077,1627)(2100,1627)(2123,1627)(2146,1627)(2170,1627)(2193,1627)(2216,1627)(2239,1627)(2262,1627)(2285,1627)(2308,1627)(2332,1627)(2355,1627)(2378,1627)(2401,1627)(2424,1627)(2447,1627)(2471,1627)(2494,1627)(2517,1627)(2540,1627)(2563,1627)(2586,1627)(2610,1627)(2633,1627)(2656,1627)(2679,1627)(2702,1627)(2725,1627)(2749,1627)
\color{green}
\thinlines \path(455,666)(455,666)(479,879)(502,1024)(525,1127)(548,1204)(571,1263)(594,1309)(618,1347)(641,1377)(664,1402)(687,1424)(710,1442)(733,1458)(757,1471)(780,1483)(803,1493)(826,1503)(849,1511)(872,1518)(896,1525)(919,1531)(942,1536)(965,1541)(988,1545)(1011,1550)(1034,1553)(1058,1557)(1081,1560)(1104,1563)(1127,1566)(1150,1568)(1173,1571)(1197,1573)(1220,1575)(1243,1577)(1266,1579)(1289,1581)(1312,1582)(1336,1584)(1359,1585)(1382,1587)(1405,1588)(1428,1589)(1451,1590)(1475,1591)(1498,1592)(1521,1593)(1544,1594)(1567,1595)(1590,1596)
\thinlines \path(1590,1596)(1614,1597)(1637,1598)(1660,1599)(1683,1599)(1706,1600)(1729,1601)(1753,1601)(1776,1602)(1799,1602)(1822,1603)(1845,1604)(1868,1604)(1892,1605)(1915,1605)(1938,1606)(1961,1606)(1984,1606)(2007,1607)(2031,1607)(2054,1608)(2077,1608)(2100,1608)(2123,1609)(2146,1609)(2170,1609)(2193,1610)(2216,1610)(2239,1610)(2262,1611)(2285,1611)(2308,1611)(2332,1612)(2355,1612)(2378,1612)(2401,1612)(2424,1613)(2447,1613)(2471,1613)(2494,1613)(2517,1613)(2540,1614)(2563,1614)(2586,1614)(2610,1614)(2633,1614)(2656,1615)(2679,1615)(2702,1615)(2725,1615)(2749,1615)
\end{picture}
\setlength{\unitlength}{0.0750pt}
\begin{picture}(3000,1800)(0,0)
\footnotesize
\color{black}
\thicklines \path(369,249)(410,249)
\thicklines \path(2876,249)(2835,249)
\put(328,249){\makebox(0,0)[r]{-2.38}}
\thicklines \path(369,459)(410,459)
\thicklines \path(2876,459)(2835,459)
\put(328,459){\makebox(0,0)[r]{-2.375}}
\thicklines \path(369,669)(410,669)
\thicklines \path(2876,669)(2835,669)
\put(328,669){\makebox(0,0)[r]{-2.37}}
\thicklines \path(369,879)(410,879)
\thicklines \path(2876,879)(2835,879)
\put(328,879){\makebox(0,0)[r]{-2.365}}
\thicklines \path(369,1088)(410,1088)
\thicklines \path(2876,1088)(2835,1088)
\put(328,1088){\makebox(0,0)[r]{-2.36}}
\thicklines \path(369,1298)(410,1298)
\thicklines \path(2876,1298)(2835,1298)
\put(328,1298){\makebox(0,0)[r]{-2.355}}
\thicklines \path(369,1508)(410,1508)
\thicklines \path(2876,1508)(2835,1508)
\put(328,1508){\makebox(0,0)[r]{-2.35}}
\thicklines \path(369,1718)(410,1718)
\thicklines \path(2876,1718)(2835,1718)
\put(328,1718){\makebox(0,0)[r]{-2.345}}
\thicklines \path(369,249)(369,290)
\thicklines \path(369,1718)(369,1677)
\put(369,166){\makebox(0,0){ 2}}
\thicklines \path(648,249)(648,290)
\thicklines \path(648,1718)(648,1677)
\put(648,166){\makebox(0,0){ 4}}
\thicklines \path(926,249)(926,290)
\thicklines \path(926,1718)(926,1677)
\put(926,166){\makebox(0,0){ 6}}
\thicklines \path(1205,249)(1205,290)
\thicklines \path(1205,1718)(1205,1677)
\put(1205,166){\makebox(0,0){ 8}}
\thicklines \path(1483,249)(1483,290)
\thicklines \path(1483,1718)(1483,1677)
\put(1483,166){\makebox(0,0){ 10}}
\thicklines \path(1762,249)(1762,290)
\thicklines \path(1762,1718)(1762,1677)
\put(1762,166){\makebox(0,0){ 12}}
\thicklines \path(2040,249)(2040,290)
\thicklines \path(2040,1718)(2040,1677)
\put(2040,166){\makebox(0,0){ 14}}
\thicklines \path(2319,249)(2319,290)
\thicklines \path(2319,1718)(2319,1677)
\put(2319,166){\makebox(0,0){ 16}}
\thicklines \path(2597,249)(2597,290)
\thicklines \path(2597,1718)(2597,1677)
\put(2597,166){\makebox(0,0){ 18}}
\thicklines \path(2876,249)(2876,290)
\thicklines \path(2876,1718)(2876,1677)
\put(2876,166){\makebox(0,0){ 20}}
\color{black}
\thicklines \path(369,249)(2876,249)(2876,1718)(369,1718)(369,249)
\put(1622,42){\makebox(0,0){$\Im\Delta\omega$}}
\color{red}
\put(648,1251){\makebox(0,0){$\Diamond$}}
\put(787,1259){\makebox(0,0){$\Diamond$}}
\put(926,1288){\makebox(0,0){$\Diamond$}}
\put(1065,1315){\makebox(0,0){$\Diamond$}}
\put(1205,1340){\makebox(0,0){$\Diamond$}}
\put(1344,1361){\makebox(0,0){$\Diamond$}}
\put(1483,1378){\makebox(0,0){$\Diamond$}}
\put(1622,1391){\makebox(0,0){$\Diamond$}}
\put(1762,1403){\makebox(0,0){$\Diamond$}}
\put(1901,1413){\makebox(0,0){$\Diamond$}}
\put(2040,1421){\makebox(0,0){$\Diamond$}}
\put(2180,1428){\makebox(0,0){$\Diamond$}}
\put(2319,1434){\makebox(0,0){$\Diamond$}}
\put(2458,1440){\makebox(0,0){$\Diamond$}}
\put(2597,1444){\makebox(0,0){$\Diamond$}}
\put(2737,1448){\makebox(0,0){$\Diamond$}}
\color{blue}
\thinlines \path(648,1508)(648,1508)(669,1508)(690,1508)(711,1508)(732,1508)(753,1508)(774,1508)(795,1508)(816,1508)(837,1508)(859,1508)(880,1508)(901,1508)(922,1508)(943,1508)(964,1508)(985,1508)(1006,1508)(1027,1508)(1049,1508)(1070,1508)(1091,1508)(1112,1508)(1133,1508)(1154,1508)(1175,1508)(1196,1508)(1217,1508)(1238,1508)(1260,1508)(1281,1508)(1302,1508)(1323,1508)(1344,1508)(1365,1508)(1386,1508)(1407,1508)(1428,1508)(1449,1508)(1471,1508)(1492,1508)(1513,1508)(1534,1508)(1555,1508)(1576,1508)(1597,1508)(1618,1508)(1639,1508)(1660,1508)(1682,1508)
\thinlines \path(1682,1508)(1703,1508)(1724,1508)(1745,1508)(1766,1508)(1787,1508)(1808,1508)(1829,1508)(1850,1508)(1872,1508)(1893,1508)(1914,1508)(1935,1508)(1956,1508)(1977,1508)(1998,1508)(2019,1508)(2040,1508)(2061,1508)(2083,1508)(2104,1508)(2125,1508)(2146,1508)(2167,1508)(2188,1508)(2209,1508)(2230,1508)(2251,1508)(2272,1508)(2294,1508)(2315,1508)(2336,1508)(2357,1508)(2378,1508)(2399,1508)(2420,1508)(2441,1508)(2462,1508)(2483,1508)(2505,1508)(2526,1508)(2547,1508)(2568,1508)(2589,1508)(2610,1508)(2631,1508)(2652,1508)(2673,1508)(2695,1508)(2716,1508)(2737,1508)
\color{green}
\thinlines \path(648,307)(648,307)(669,372)(690,431)(711,486)(732,536)(753,582)(774,624)(795,664)(816,700)(837,734)(859,766)(880,796)(901,823)(922,849)(943,873)(964,896)(985,918)(1006,938)(1027,957)(1049,975)(1070,993)(1091,1009)(1112,1024)(1133,1039)(1154,1053)(1175,1066)(1196,1078)(1217,1091)(1238,1102)(1260,1113)(1281,1123)(1302,1133)(1323,1143)(1344,1152)(1365,1161)(1386,1169)(1407,1178)(1428,1185)(1449,1193)(1471,1200)(1492,1207)(1513,1214)(1534,1220)(1555,1226)(1576,1232)(1597,1238)(1618,1244)(1639,1249)(1660,1254)(1682,1259)
\thinlines \path(1682,1259)(1703,1264)(1724,1269)(1745,1273)(1766,1278)(1787,1282)(1808,1286)(1829,1290)(1850,1294)(1872,1298)(1893,1302)(1914,1305)(1935,1309)(1956,1312)(1977,1315)(1998,1319)(2019,1322)(2040,1325)(2061,1328)(2083,1330)(2104,1333)(2125,1336)(2146,1339)(2167,1341)(2188,1344)(2209,1346)(2230,1349)(2251,1351)(2272,1353)(2294,1355)(2315,1358)(2336,1360)(2357,1362)(2378,1364)(2399,1366)(2420,1368)(2441,1370)(2462,1371)(2483,1373)(2505,1375)(2526,1377)(2547,1378)(2568,1380)(2589,1382)(2610,1383)(2631,1385)(2652,1386)(2673,1388)(2695,1389)(2716,1391)(2737,1392)
\end{picture}
\caption{\label{fig1}The frequency gap~(\ref{eqgap}) for scalar perturbations in $d=4$ for $r_H=1$ and $\ell = 2$: zeroth and first order analytical (eq.~(\ref{eq109})) compared with numerical data~\cite{CKL}.}
\end{center}
\end{figure}
For scalar perturbations, we find from eqs.~(\ref{eqo1st}), (\ref{eqab}),
together with~(\ref{eqVS0}), (\ref{eqVS0a}) and (\ref{eqV0}),
\be \o_n \bar r_* = \left(n + \frac{1}{4} \right)\pi + \frac{i}{2}\ln 2 +
e^{i\pi /4} \frac{\mathcal{A}_{\mathsf{S}}\Gamma^4 (\4 )}{16\pi^2}
\sqrt{\frac{\bar r_*}{2\mu n}} \ , \ \ \ \ \mathcal{A}_{\mathsf{S}} = \frac{\ell(\ell +1) -1}{6} \ee
Notice that only the first-order correction is $\ell$-dependent.
In the limit of large horizon radius ($r_H \approx (2\mu)^{1/3} \gg 1$), we have
from~(\ref{eqrs})
\be\label{eq107} \bar r_* \approx \frac{\pi (1+i\sqrt 3)}{3\sqrt 3 r_H} \ee
Numerically for $\ell =2$,
\be \frac{\o_n}{r_H} = (1.299-2.250i)n +0.573-0.419i +\frac{0.508+0.293i}{r_H^2 \sqrt n} \ee
For an intermediate black hole, $r_H = 1$, we obtain
\be\label{eq109} \o_n = (1.969-2.350i)n + 0.752-0.370i +\frac{0.654+0.458i}{\sqrt n} \ee
In figure~\ref{fig1} we compare this analytical result with numerical results~\cite{CKL}.
We plot the gap
\be\label{eqgap} \Delta\omega_n = \omega_n-\omega_{n-1} \ee
because the offset does not always agree with numerical results~\cite{NS}.
We show both zeroth-order and first-order analytical results.
For a small black hole, $r_H = 0.2$, we obtain
\be \o_n = (1.695-0.571i)n + 0.487-0.0441i +\frac{1.093+0.561i}{\sqrt n} \ee
\begin{figure}
\begin{center}
\setlength{\unitlength}{0.0750pt}
\begin{picture}(3000,1800)(0,0)
\footnotesize
\color{black}
\thicklines \path(369,249)(410,249)
\thicklines \path(2876,249)(2835,249)
\put(328,249){\makebox(0,0)[r]{ 1.963}}
\thicklines \path(369,459)(410,459)
\thicklines \path(2876,459)(2835,459)
\put(328,459){\makebox(0,0)[r]{ 1.964}}
\thicklines \path(369,669)(410,669)
\thicklines \path(2876,669)(2835,669)
\put(328,669){\makebox(0,0)[r]{ 1.965}}
\thicklines \path(369,879)(410,879)
\thicklines \path(2876,879)(2835,879)
\put(328,879){\makebox(0,0)[r]{ 1.966}}
\thicklines \path(369,1088)(410,1088)
\thicklines \path(2876,1088)(2835,1088)
\put(328,1088){\makebox(0,0)[r]{ 1.967}}
\thicklines \path(369,1298)(410,1298)
\thicklines \path(2876,1298)(2835,1298)
\put(328,1298){\makebox(0,0)[r]{ 1.968}}
\thicklines \path(369,1508)(410,1508)
\thicklines \path(2876,1508)(2835,1508)
\put(328,1508){\makebox(0,0)[r]{ 1.969}}
\thicklines \path(369,1718)(410,1718)
\thicklines \path(2876,1718)(2835,1718)
\put(328,1718){\makebox(0,0)[r]{ 1.97}}
\thicklines \path(369,249)(369,290)
\thicklines \path(369,1718)(369,1677)
\put(369,166){\makebox(0,0){ 4}}
\thicklines \path(682,249)(682,290)
\thicklines \path(682,1718)(682,1677)
\put(682,166){\makebox(0,0){ 6}}
\thicklines \path(996,249)(996,290)
\thicklines \path(996,1718)(996,1677)
\put(996,166){\makebox(0,0){ 8}}
\thicklines \path(1309,249)(1309,290)
\thicklines \path(1309,1718)(1309,1677)
\put(1309,166){\makebox(0,0){ 10}}
\thicklines \path(1623,249)(1623,290)
\thicklines \path(1623,1718)(1623,1677)
\put(1623,166){\makebox(0,0){ 12}}
\thicklines \path(1936,249)(1936,290)
\thicklines \path(1936,1718)(1936,1677)
\put(1936,166){\makebox(0,0){ 14}}
\thicklines \path(2249,249)(2249,290)
\thicklines \path(2249,1718)(2249,1677)
\put(2249,166){\makebox(0,0){ 16}}
\thicklines \path(2563,249)(2563,290)
\thicklines \path(2563,1718)(2563,1677)
\put(2563,166){\makebox(0,0){ 18}}
\thicklines \path(2876,249)(2876,290)
\thicklines \path(2876,1718)(2876,1677)
\put(2876,166){\makebox(0,0){ 20}}
\color{black}
\thicklines \path(369,249)(2876,249)(2876,1718)(369,1718)(369,249)
\put(1622,42){\makebox(0,0){$\Re\Delta\omega$}}
\color{red}
\put(526,753){\makebox(0,0){$\Diamond$}}
\put(682,900){\makebox(0,0){$\Diamond$}}
\put(839,1025){\makebox(0,0){$\Diamond$}}
\put(996,1088){\makebox(0,0){$\Diamond$}}
\put(1152,1151){\makebox(0,0){$\Diamond$}}
\put(1309,1195){\makebox(0,0){$\Diamond$}}
\put(1466,1240){\makebox(0,0){$\Diamond$}}
\put(1623,1271){\makebox(0,0){$\Diamond$}}
\put(1779,1296){\makebox(0,0){$\Diamond$}}
\put(1936,1319){\makebox(0,0){$\Diamond$}}
\put(2093,1338){\makebox(0,0){$\Diamond$}}
\put(2249,1355){\makebox(0,0){$\Diamond$}}
\put(2406,1368){\makebox(0,0){$\Diamond$}}
\put(2563,1380){\makebox(0,0){$\Diamond$}}
\put(2719,1393){\makebox(0,0){$\Diamond$}}
\color{blue}
\thinlines \path(526,1508)(526,1508)(548,1508)(570,1508)(592,1508)(614,1508)(636,1508)(659,1508)(681,1508)(703,1508)(725,1508)(747,1508)(769,1508)(792,1508)(814,1508)(836,1508)(858,1508)(880,1508)(902,1508)(925,1508)(947,1508)(969,1508)(991,1508)(1013,1508)(1035,1508)(1057,1508)(1080,1508)(1102,1508)(1124,1508)(1146,1508)(1168,1508)(1190,1508)(1213,1508)(1235,1508)(1257,1508)(1279,1508)(1301,1508)(1323,1508)(1346,1508)(1368,1508)(1390,1508)(1412,1508)(1434,1508)(1456,1508)(1478,1508)(1501,1508)(1523,1508)(1545,1508)(1567,1508)(1589,1508)(1611,1508)
\thinlines \path(1611,1508)(1634,1508)(1656,1508)(1678,1508)(1700,1508)(1722,1508)(1744,1508)(1767,1508)(1789,1508)(1811,1508)(1833,1508)(1855,1508)(1877,1508)(1899,1508)(1922,1508)(1944,1508)(1966,1508)(1988,1508)(2010,1508)(2032,1508)(2055,1508)(2077,1508)(2099,1508)(2121,1508)(2143,1508)(2165,1508)(2188,1508)(2210,1508)(2232,1508)(2254,1508)(2276,1508)(2298,1508)(2320,1508)(2343,1508)(2365,1508)(2387,1508)(2409,1508)(2431,1508)(2453,1508)(2476,1508)(2498,1508)(2520,1508)(2542,1508)(2564,1508)(2586,1508)(2609,1508)(2631,1508)(2653,1508)(2675,1508)(2697,1508)(2719,1508)
\color{green}
\thinlines \path(526,279)(526,279)(548,329)(570,376)(592,420)(614,461)(636,500)(659,536)(681,571)(703,603)(725,633)(747,662)(769,689)(792,715)(814,739)(836,763)(858,785)(880,806)(902,826)(925,845)(947,863)(969,881)(991,897)(1013,913)(1035,928)(1057,943)(1080,957)(1102,970)(1124,983)(1146,996)(1168,1007)(1190,1019)(1213,1030)(1235,1041)(1257,1051)(1279,1061)(1301,1070)(1323,1079)(1346,1088)(1368,1097)(1390,1105)(1412,1113)(1434,1121)(1456,1128)(1478,1135)(1501,1143)(1523,1149)(1545,1156)(1567,1162)(1589,1169)(1611,1175)
\thinlines \path(1611,1175)(1634,1180)(1656,1186)(1678,1192)(1700,1197)(1722,1202)(1744,1207)(1767,1212)(1789,1217)(1811,1222)(1833,1226)(1855,1231)(1877,1235)(1899,1239)(1922,1243)(1944,1247)(1966,1251)(1988,1255)(2010,1259)(2032,1262)(2055,1266)(2077,1269)(2099,1272)(2121,1276)(2143,1279)(2165,1282)(2188,1285)(2210,1288)(2232,1291)(2254,1294)(2276,1297)(2298,1300)(2320,1302)(2343,1305)(2365,1307)(2387,1310)(2409,1312)(2431,1315)(2453,1317)(2476,1319)(2498,1322)(2520,1324)(2542,1326)(2564,1328)(2586,1330)(2609,1332)(2631,1334)(2653,1336)(2675,1338)(2697,1340)(2719,1342)
\end{picture}
\setlength{\unitlength}{0.0750pt}
\begin{picture}(3000,1800)(0,0)
\footnotesize
\color{black}
\thicklines \path(328,249)(369,249)
\thicklines \path(2876,249)(2835,249)
\put(287,249){\makebox(0,0)[r]{-2.4}}
\thicklines \path(328,516)(369,516)
\thicklines \path(2876,516)(2835,516)
\put(287,516){\makebox(0,0)[r]{-2.39}}
\thicklines \path(328,783)(369,783)
\thicklines \path(2876,783)(2835,783)
\put(287,783){\makebox(0,0)[r]{-2.38}}
\thicklines \path(328,1050)(369,1050)
\thicklines \path(2876,1050)(2835,1050)
\put(287,1050){\makebox(0,0)[r]{-2.37}}
\thicklines \path(328,1317)(369,1317)
\thicklines \path(2876,1317)(2835,1317)
\put(287,1317){\makebox(0,0)[r]{-2.36}}
\thicklines \path(328,1584)(369,1584)
\thicklines \path(2876,1584)(2835,1584)
\put(287,1584){\makebox(0,0)[r]{-2.35}}
\thicklines \path(328,249)(328,290)
\thicklines \path(328,1718)(328,1677)
\put(328,166){\makebox(0,0){ 0}}
\thicklines \path(583,249)(583,290)
\thicklines \path(583,1718)(583,1677)
\put(583,166){\makebox(0,0){ 2}}
\thicklines \path(838,249)(838,290)
\thicklines \path(838,1718)(838,1677)
\put(838,166){\makebox(0,0){ 4}}
\thicklines \path(1092,249)(1092,290)
\thicklines \path(1092,1718)(1092,1677)
\put(1092,166){\makebox(0,0){ 6}}
\thicklines \path(1347,249)(1347,290)
\thicklines \path(1347,1718)(1347,1677)
\put(1347,166){\makebox(0,0){ 8}}
\thicklines \path(1602,249)(1602,290)
\thicklines \path(1602,1718)(1602,1677)
\put(1602,166){\makebox(0,0){ 10}}
\thicklines \path(1857,249)(1857,290)
\thicklines \path(1857,1718)(1857,1677)
\put(1857,166){\makebox(0,0){ 12}}
\thicklines \path(2112,249)(2112,290)
\thicklines \path(2112,1718)(2112,1677)
\put(2112,166){\makebox(0,0){ 14}}
\thicklines \path(2366,249)(2366,290)
\thicklines \path(2366,1718)(2366,1677)
\put(2366,166){\makebox(0,0){ 16}}
\thicklines \path(2621,249)(2621,290)
\thicklines \path(2621,1718)(2621,1677)
\put(2621,166){\makebox(0,0){ 18}}
\thicklines \path(2876,249)(2876,290)
\thicklines \path(2876,1718)(2876,1677)
\put(2876,166){\makebox(0,0){ 20}}
\color{black}
\thicklines \path(328,249)(2876,249)(2876,1718)(328,1718)(328,249)
\put(1602,42){\makebox(0,0){$\Im\Delta\omega$}}
\color{red}
\put(455,1147){\makebox(0,0){$\Diamond$}}
\put(583,1400){\makebox(0,0){$\Diamond$}}
\put(710,1477){\makebox(0,0){$\Diamond$}}
\put(838,1513){\makebox(0,0){$\Diamond$}}
\put(965,1528){\makebox(0,0){$\Diamond$}}
\put(1092,1542){\makebox(0,0){$\Diamond$}}
\put(1220,1550){\makebox(0,0){$\Diamond$}}
\put(1347,1555){\makebox(0,0){$\Diamond$}}
\put(1475,1558){\makebox(0,0){$\Diamond$}}
\put(1602,1562){\makebox(0,0){$\Diamond$}}
\put(1729,1564){\makebox(0,0){$\Diamond$}}
\put(1857,1566){\makebox(0,0){$\Diamond$}}
\put(1984,1567){\makebox(0,0){$\Diamond$}}
\put(2112,1568){\makebox(0,0){$\Diamond$}}
\put(2239,1570){\makebox(0,0){$\Diamond$}}
\put(2366,1571){\makebox(0,0){$\Diamond$}}
\put(2494,1571){\makebox(0,0){$\Diamond$}}
\put(2621,1572){\makebox(0,0){$\Diamond$}}
\put(2749,1573){\makebox(0,0){$\Diamond$}}
\color{blue}
\thinlines \path(455,1584)(455,1584)(479,1584)(502,1584)(525,1584)(548,1584)(571,1584)(594,1584)(618,1584)(641,1584)(664,1584)(687,1584)(710,1584)(733,1584)(757,1584)(780,1584)(803,1584)(826,1584)(849,1584)(872,1584)(896,1584)(919,1584)(942,1584)(965,1584)(988,1584)(1011,1584)(1034,1584)(1058,1584)(1081,1584)(1104,1584)(1127,1584)(1150,1584)(1173,1584)(1197,1584)(1220,1584)(1243,1584)(1266,1584)(1289,1584)(1312,1584)(1336,1584)(1359,1584)(1382,1584)(1405,1584)(1428,1584)(1451,1584)(1475,1584)(1498,1584)(1521,1584)(1544,1584)(1567,1584)(1590,1584)
\thinlines \path(1590,1584)(1614,1584)(1637,1584)(1660,1584)(1683,1584)(1706,1584)(1729,1584)(1753,1584)(1776,1584)(1799,1584)(1822,1584)(1845,1584)(1868,1584)(1892,1584)(1915,1584)(1938,1584)(1961,1584)(1984,1584)(2007,1584)(2031,1584)(2054,1584)(2077,1584)(2100,1584)(2123,1584)(2146,1584)(2170,1584)(2193,1584)(2216,1584)(2239,1584)(2262,1584)(2285,1584)(2308,1584)(2332,1584)(2355,1584)(2378,1584)(2401,1584)(2424,1584)(2447,1584)(2471,1584)(2494,1584)(2517,1584)(2540,1584)(2563,1584)(2586,1584)(2610,1584)(2633,1584)(2656,1584)(2679,1584)(2702,1584)(2725,1584)(2749,1584)
\color{green}
\thinlines \path(455,361)(455,361)(479,632)(502,816)(525,948)(548,1046)(571,1121)(594,1180)(618,1227)(641,1266)(664,1299)(687,1326)(710,1349)(733,1369)(757,1386)(780,1401)(803,1414)(826,1426)(849,1437)(872,1446)(896,1454)(919,1462)(942,1469)(965,1475)(988,1481)(1011,1486)(1034,1491)(1058,1495)(1081,1499)(1104,1503)(1127,1507)(1150,1510)(1173,1513)(1197,1516)(1220,1518)(1243,1521)(1266,1523)(1289,1525)(1312,1528)(1336,1529)(1359,1531)(1382,1533)(1405,1535)(1428,1536)(1451,1538)(1475,1539)(1498,1540)(1521,1542)(1544,1543)(1567,1544)(1590,1545)
\thinlines \path(1590,1545)(1614,1546)(1637,1547)(1660,1548)(1683,1549)(1706,1550)(1729,1551)(1753,1552)(1776,1553)(1799,1553)(1822,1554)(1845,1555)(1868,1555)(1892,1556)(1915,1557)(1938,1557)(1961,1558)(1984,1558)(2007,1559)(2031,1559)(2054,1560)(2077,1560)(2100,1561)(2123,1561)(2146,1562)(2170,1562)(2193,1563)(2216,1563)(2239,1563)(2262,1564)(2285,1564)(2308,1564)(2332,1565)(2355,1565)(2378,1566)(2401,1566)(2424,1566)(2447,1566)(2471,1567)(2494,1567)(2517,1567)(2540,1568)(2563,1568)(2586,1568)(2610,1568)(2633,1569)(2656,1569)(2679,1569)(2702,1569)(2725,1569)(2749,1570)
\end{picture}
\caption{\label{fig2}The frequency gap~(\ref{eqgap}) for tensor perturbations in $d=4$ for $r_H=1$ and $\ell = 0$: zeroth and first order analytical (eq.~(\ref{eq114})) compared with numerical data~\cite{CKL}.}
\end{center}
\end{figure}
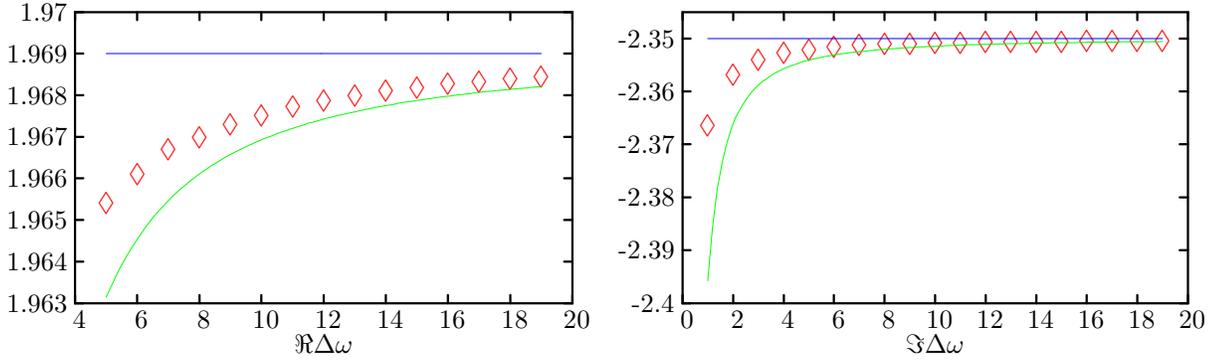
For tensor perturbations, we find from eqs.~(\ref{eqo1st}), (\ref{eqab}),
together with~(\ref{eqVT0}) and (\ref{eqV0}),
\be \o_n \bar r_* = \left(n + \frac{1}{4} \right)\pi + \frac{i}{2}\ln 2 +
e^{i\pi /4} \frac{\mathcal{A}_{\mathsf{T}}\Gamma^4 (\4 )}{16\pi^2}
\sqrt{\frac{\bar r_*}{2\mu n}} \ , \ \ \ \ \mathcal{A}_{\mathsf{T}} = \frac{3\ell(\ell +1) +1}{6} \ee
Again, only the first-order correction is $\ell$-dependent.
Numerically for large $r_H$ and $\ell =0$,
\be \frac{\o_n}{r_H} = (1.299-2.250i)n +0.573-0.419i +\frac{0.102+0.0586i}{r_H^2 \sqrt n} \ee
For an intermediate black hole, $r_H = 1$, we obtain
\be\label{eq114} \o_n = (1.969-2.350i)n + 0.752-0.370i +\frac{0.131+0.0916i}{\sqrt n} \ee
In figure~\ref{fig2}, we plot the gap~(\ref{eqgap}), including both zeroth and first order
and compare with numerical results~\cite{CKL}.

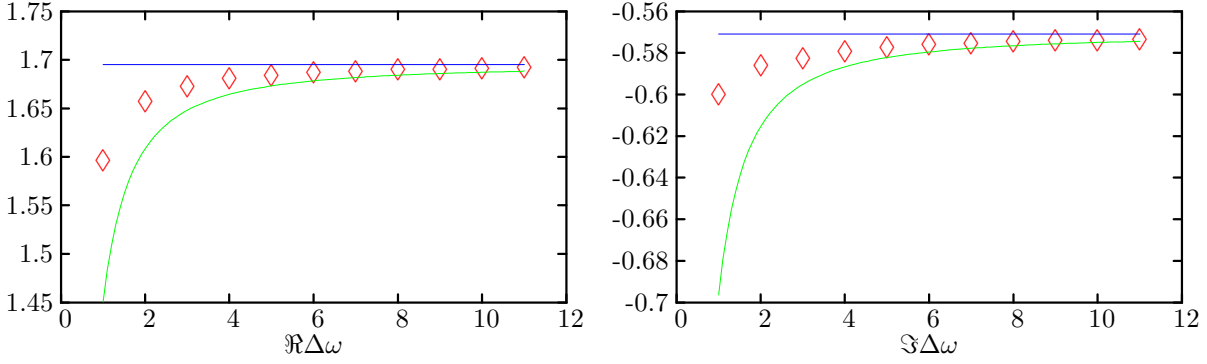
\begin{figure}
\begin{center}
\setlength{\unitlength}{0.0750pt}
\begin{picture}(3000,1800)(0,0)
\footnotesize
\color{black}
\thicklines \path(328,249)(369,249)
\thicklines \path(2876,249)(2835,249)
\put(287,249){\makebox(0,0)[r]{ 1.45}}
\thicklines \path(328,494)(369,494)
\thicklines \path(2876,494)(2835,494)
\put(287,494){\makebox(0,0)[r]{ 1.5}}
\thicklines \path(328,739)(369,739)
\thicklines \path(2876,739)(2835,739)
\put(287,739){\makebox(0,0)[r]{ 1.55}}
\thicklines \path(328,984)(369,984)
\thicklines \path(2876,984)(2835,984)
\put(287,984){\makebox(0,0)[r]{ 1.6}}
\thicklines \path(328,1228)(369,1228)
\thicklines \path(2876,1228)(2835,1228)
\put(287,1228){\makebox(0,0)[r]{ 1.65}}
\thicklines \path(328,1473)(369,1473)
\thicklines \path(2876,1473)(2835,1473)
\put(287,1473){\makebox(0,0)[r]{ 1.7}}
\thicklines \path(328,1718)(369,1718)
\thicklines \path(2876,1718)(2835,1718)
\put(287,1718){\makebox(0,0)[r]{ 1.75}}
\thicklines \path(328,249)(328,290)
\thicklines \path(328,1718)(328,1677)
\put(328,166){\makebox(0,0){ 0}}
\thicklines \path(753,249)(753,290)
\thicklines \path(753,1718)(753,1677)
\put(753,166){\makebox(0,0){ 2}}
\thicklines \path(1177,249)(1177,290)
\thicklines \path(1177,1718)(1177,1677)
\put(1177,166){\makebox(0,0){ 4}}
\thicklines \path(1602,249)(1602,290)
\thicklines \path(1602,1718)(1602,1677)
\put(1602,166){\makebox(0,0){ 6}}
\thicklines \path(2027,249)(2027,290)
\thicklines \path(2027,1718)(2027,1677)
\put(2027,166){\makebox(0,0){ 8}}
\thicklines \path(2451,249)(2451,290)
\thicklines \path(2451,1718)(2451,1677)
\put(2451,166){\makebox(0,0){ 10}}
\thicklines \path(2876,249)(2876,290)
\thicklines \path(2876,1718)(2876,1677)
\put(2876,166){\makebox(0,0){ 12}}
\color{black}
\thicklines \path(328,249)(2876,249)(2876,1718)(328,1718)(328,249)
\put(1602,42){\makebox(0,0){$\Re\Delta\omega$}}
\color{red}
\put(540,963){\makebox(0,0){$\Diamond$}}
\put(753,1262){\makebox(0,0){$\Diamond$}}
\put(965,1341){\makebox(0,0){$\Diamond$}}
\put(1177,1377){\makebox(0,0){$\Diamond$}}
\put(1390,1396){\makebox(0,0){$\Diamond$}}
\put(1602,1408){\makebox(0,0){$\Diamond$}}
\put(1814,1417){\makebox(0,0){$\Diamond$}}
\put(2027,1422){\makebox(0,0){$\Diamond$}}
\put(2239,1426){\makebox(0,0){$\Diamond$}}
\put(2451,1430){\makebox(0,0){$\Diamond$}}
\put(2664,1432){\makebox(0,0){$\Diamond$}}
\color{blue}
\thinlines \path(540,1449)(540,1449)(562,1449)(583,1449)(605,1449)(626,1449)(648,1449)(669,1449)(690,1449)(712,1449)(733,1449)(755,1449)(776,1449)(798,1449)(819,1449)(841,1449)(862,1449)(883,1449)(905,1449)(926,1449)(948,1449)(969,1449)(991,1449)(1012,1449)(1034,1449)(1055,1449)(1077,1449)(1098,1449)(1119,1449)(1141,1449)(1162,1449)(1184,1449)(1205,1449)(1227,1449)(1248,1449)(1270,1449)(1291,1449)(1312,1449)(1334,1449)(1355,1449)(1377,1449)(1398,1449)(1420,1449)(1441,1449)(1463,1449)(1484,1449)(1505,1449)(1527,1449)(1548,1449)(1570,1449)(1591,1449)
\thinlines \path(1591,1449)(1613,1449)(1634,1449)(1656,1449)(1677,1449)(1699,1449)(1720,1449)(1741,1449)(1763,1449)(1784,1449)(1806,1449)(1827,1449)(1849,1449)(1870,1449)(1892,1449)(1913,1449)(1934,1449)(1956,1449)(1977,1449)(1999,1449)(2020,1449)(2042,1449)(2063,1449)(2085,1449)(2106,1449)(2127,1449)(2149,1449)(2170,1449)(2192,1449)(2213,1449)(2235,1449)(2256,1449)(2278,1449)(2299,1449)(2321,1449)(2342,1449)(2363,1449)(2385,1449)(2406,1449)(2428,1449)(2449,1449)(2471,1449)(2492,1449)(2514,1449)(2535,1449)(2556,1449)(2578,1449)(2599,1449)(2621,1449)(2642,1449)(2664,1449)
\color{green}
\thinlines \path(540,251)(540,251)(562,412)(583,540)(605,644)(626,729)(648,800)(669,860)(690,912)(712,956)(733,995)(755,1029)(776,1058)(798,1085)(819,1108)(841,1130)(862,1149)(883,1166)(905,1181)(926,1196)(948,1209)(969,1221)(991,1232)(1012,1242)(1034,1251)(1055,1260)(1077,1268)(1098,1275)(1119,1282)(1141,1289)(1162,1295)(1184,1301)(1205,1306)(1227,1311)(1248,1316)(1270,1320)(1291,1325)(1312,1329)(1334,1333)(1355,1336)(1377,1340)(1398,1343)(1420,1346)(1441,1349)(1463,1352)(1484,1354)(1505,1357)(1527,1359)(1548,1362)(1570,1364)(1591,1366)
\thinlines \path(1591,1366)(1613,1368)(1634,1370)(1656,1372)(1677,1374)(1699,1376)(1720,1377)(1741,1379)(1763,1381)(1784,1382)(1806,1383)(1827,1385)(1849,1386)(1870,1388)(1892,1389)(1913,1390)(1934,1391)(1956,1392)(1977,1393)(1999,1394)(2020,1395)(2042,1396)(2063,1397)(2085,1398)(2106,1399)(2127,1400)(2149,1401)(2170,1402)(2192,1403)(2213,1403)(2235,1404)(2256,1405)(2278,1406)(2299,1406)(2321,1407)(2342,1408)(2363,1408)(2385,1409)(2406,1410)(2428,1410)(2449,1411)(2471,1411)(2492,1412)(2514,1412)(2535,1413)(2556,1413)(2578,1414)(2599,1414)(2621,1415)(2642,1415)(2664,1416)
\end{picture}
\setlength{\unitlength}{0.0750pt}
\begin{picture}(3000,1800)(0,0)
\footnotesize
\color{black}
\thicklines \path(328,249)(369,249)
\thicklines \path(2876,249)(2835,249)
\put(287,249){\makebox(0,0)[r]{-0.7}}
\thicklines \path(328,459)(369,459)
\thicklines \path(2876,459)(2835,459)
\put(287,459){\makebox(0,0)[r]{-0.68}}
\thicklines \path(328,669)(369,669)
\thicklines \path(2876,669)(2835,669)
\put(287,669){\makebox(0,0)[r]{-0.66}}
\thicklines \path(328,879)(369,879)
\thicklines \path(2876,879)(2835,879)
\put(287,879){\makebox(0,0)[r]{-0.64}}
\thicklines \path(328,1088)(369,1088)
\thicklines \path(2876,1088)(2835,1088)
\put(287,1088){\makebox(0,0)[r]{-0.62}}
\thicklines \path(328,1298)(369,1298)
\thicklines \path(2876,1298)(2835,1298)
\put(287,1298){\makebox(0,0)[r]{-0.6}}
\thicklines \path(328,1508)(369,1508)
\thicklines \path(2876,1508)(2835,1508)
\put(287,1508){\makebox(0,0)[r]{-0.58}}
\thicklines \path(328,1718)(369,1718)
\thicklines \path(2876,1718)(2835,1718)
\put(287,1718){\makebox(0,0)[r]{-0.56}}
\thicklines \path(328,249)(328,290)
\thicklines \path(328,1718)(328,1677)
\put(328,166){\makebox(0,0){ 0}}
\thicklines \path(753,249)(753,290)
\thicklines \path(753,1718)(753,1677)
\put(753,166){\makebox(0,0){ 2}}
\thicklines \path(1177,249)(1177,290)
\thicklines \path(1177,1718)(1177,1677)
\put(1177,166){\makebox(0,0){ 4}}
\thicklines \path(1602,249)(1602,290)
\thicklines \path(1602,1718)(1602,1677)
\put(1602,166){\makebox(0,0){ 6}}
\thicklines \path(2027,249)(2027,290)
\thicklines \path(2027,1718)(2027,1677)
\put(2027,166){\makebox(0,0){ 8}}
\thicklines \path(2451,249)(2451,290)
\thicklines \path(2451,1718)(2451,1677)
\put(2451,166){\makebox(0,0){ 10}}
\thicklines \path(2876,249)(2876,290)
\thicklines \path(2876,1718)(2876,1677)
\put(2876,166){\makebox(0,0){ 12}}
\color{black}
\thicklines \path(328,249)(2876,249)(2876,1718)(328,1718)(328,249)
\put(1602,42){\makebox(0,0){$\Im\Delta\omega$}}
\color{red}
\put(540,1301){\makebox(0,0){$\Diamond$}}
\put(753,1442){\makebox(0,0){$\Diamond$}}
\put(965,1480){\makebox(0,0){$\Diamond$}}
\put(1177,1514){\makebox(0,0){$\Diamond$}}
\put(1390,1535){\makebox(0,0){$\Diamond$}}
\put(1602,1548){\makebox(0,0){$\Diamond$}}
\put(1814,1557){\makebox(0,0){$\Diamond$}}
\put(2027,1564){\makebox(0,0){$\Diamond$}}
\put(2239,1569){\makebox(0,0){$\Diamond$}}
\put(2451,1572){\makebox(0,0){$\Diamond$}}
\put(2664,1576){\makebox(0,0){$\Diamond$}}
\color{blue}
\thinlines \path(540,1603)(540,1603)(562,1603)(583,1603)(605,1603)(626,1603)(648,1603)(669,1603)(690,1603)(712,1603)(733,1603)(755,1603)(776,1603)(798,1603)(819,1603)(841,1603)(862,1603)(883,1603)(905,1603)(926,1603)(948,1603)(969,1603)(991,1603)(1012,1603)(1034,1603)(1055,1603)(1077,1603)(1098,1603)(1119,1603)(1141,1603)(1162,1603)(1184,1603)(1205,1603)(1227,1603)(1248,1603)(1270,1603)(1291,1603)(1312,1603)(1334,1603)(1355,1603)(1377,1603)(1398,1603)(1420,1603)(1441,1603)(1463,1603)(1484,1603)(1505,1603)(1527,1603)(1548,1603)(1570,1603)(1591,1603)
\thinlines \path(1591,1603)(1613,1603)(1634,1603)(1656,1603)(1677,1603)(1699,1603)(1720,1603)(1741,1603)(1763,1603)(1784,1603)(1806,1603)(1827,1603)(1849,1603)(1870,1603)(1892,1603)(1913,1603)(1934,1603)(1956,1603)(1977,1603)(1999,1603)(2020,1603)(2042,1603)(2063,1603)(2085,1603)(2106,1603)(2127,1603)(2149,1603)(2170,1603)(2192,1603)(2213,1603)(2235,1603)(2256,1603)(2278,1603)(2299,1603)(2321,1603)(2342,1603)(2363,1603)(2385,1603)(2406,1603)(2428,1603)(2449,1603)(2471,1603)(2492,1603)(2514,1603)(2535,1603)(2556,1603)(2578,1603)(2599,1603)(2621,1603)(2642,1603)(2664,1603)
\color{green}
\thinlines \path(540,286)(540,286)(562,463)(583,603)(605,717)(626,811)(648,889)(669,956)(690,1012)(712,1061)(733,1103)(755,1141)(776,1173)(798,1202)(819,1228)(841,1252)(862,1272)(883,1291)(905,1309)(926,1324)(948,1339)(969,1352)(991,1364)(1012,1375)(1034,1385)(1055,1395)(1077,1404)(1098,1412)(1119,1420)(1141,1427)(1162,1434)(1184,1440)(1205,1446)(1227,1451)(1248,1457)(1270,1462)(1291,1466)(1312,1471)(1334,1475)(1355,1479)(1377,1483)(1398,1486)(1420,1490)(1441,1493)(1463,1496)(1484,1499)(1505,1502)(1527,1504)(1548,1507)(1570,1509)(1591,1512)
\thinlines \path(1591,1512)(1613,1514)(1634,1516)(1656,1518)(1677,1520)(1699,1522)(1720,1524)(1741,1526)(1763,1528)(1784,1529)(1806,1531)(1827,1532)(1849,1534)(1870,1535)(1892,1537)(1913,1538)(1934,1539)(1956,1541)(1977,1542)(1999,1543)(2020,1544)(2042,1545)(2063,1546)(2085,1547)(2106,1548)(2127,1549)(2149,1550)(2170,1551)(2192,1552)(2213,1553)(2235,1554)(2256,1554)(2278,1555)(2299,1556)(2321,1557)(2342,1557)(2363,1558)(2385,1559)(2406,1560)(2428,1560)(2449,1561)(2471,1561)(2492,1562)(2514,1563)(2535,1563)(2556,1564)(2578,1564)(2599,1565)(2621,1565)(2642,1566)(2664,1566)
\end{picture}
\caption{\label{fig3}The frequency gap~(\ref{eqgap}) for tensor perturbations in $d=4$ for $r_H=0.2$ and $\ell = 0$: zeroth and first order analytical (eq.~(\ref{eq115})) compared with numerical data~\cite{CKL}.}
\end{center}
\end{figure}

For a small black hole, $r_H = 0.2$, we obtain
\be\label{eq115} \o_n = (1.695-0.571i)n + 0.487-0.0441i +\frac{0.489+0.251i}{\sqrt n} \ee
and compare the gap with numerical results in figure~\ref{fig3}.
Finally, for vector perturbations, we find from eqs.~(\ref{eqo1st}), (\ref{eqab}),
together with~(\ref{eqVV0}) and (\ref{eqV0}),
\be \o_n \bar r_* = \left(n + \frac{1}{4} \right)\pi + \frac{i}{2}\ln 2 +
e^{i\pi /4} \frac{\mathcal{A}_{\mathsf{V}}\Gamma^4 (\4 )}{48\pi^2}
\sqrt{\frac{\bar r_*}{2\mu n}} \ , \ \ \ \ \mathcal{A}_{\mathsf{V}} = \frac{\ell(\ell +1)}{2} + \frac{3}{14} \ee
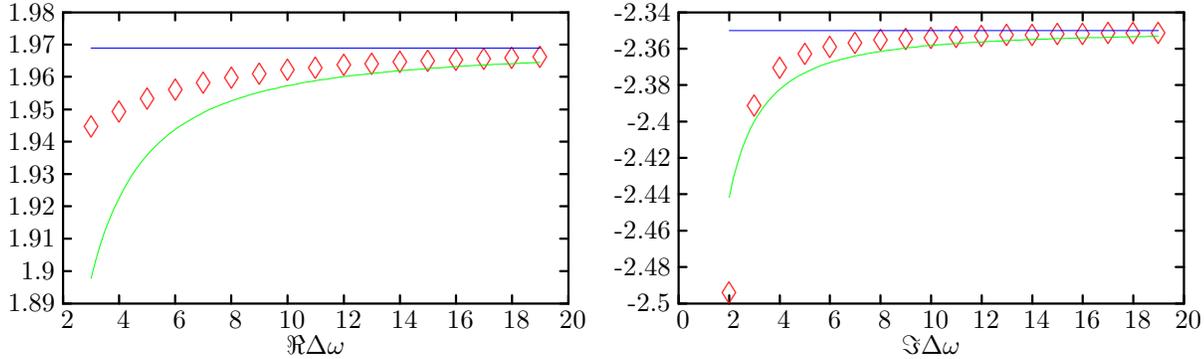
\begin{figure}[ht]
\begin{center}
\setlength{\unitlength}{0.0750pt}
\begin{picture}(3000,1800)(0,0)
\footnotesize
\color{black}
\thicklines \path(328,249)(369,249)
\thicklines \path(2876,249)(2835,249)
\put(287,249){\makebox(0,0)[r]{ 1.89}}
\thicklines \path(328,412)(369,412)
\thicklines \path(2876,412)(2835,412)
\put(287,412){\makebox(0,0)[r]{ 1.9}}
\thicklines \path(328,575)(369,575)
\thicklines \path(2876,575)(2835,575)
\put(287,575){\makebox(0,0)[r]{ 1.91}}
\thicklines \path(328,739)(369,739)
\thicklines \path(2876,739)(2835,739)
\put(287,739){\makebox(0,0)[r]{ 1.92}}
\thicklines \path(328,902)(369,902)
\thicklines \path(2876,902)(2835,902)
\put(287,902){\makebox(0,0)[r]{ 1.93}}
\thicklines \path(328,1065)(369,1065)
\thicklines \path(2876,1065)(2835,1065)
\put(287,1065){\makebox(0,0)[r]{ 1.94}}
\thicklines \path(328,1228)(369,1228)
\thicklines \path(2876,1228)(2835,1228)
\put(287,1228){\makebox(0,0)[r]{ 1.95}}
\thicklines \path(328,1392)(369,1392)
\thicklines \path(2876,1392)(2835,1392)
\put(287,1392){\makebox(0,0)[r]{ 1.96}}
\thicklines \path(328,1555)(369,1555)
\thicklines \path(2876,1555)(2835,1555)
\put(287,1555){\makebox(0,0)[r]{ 1.97}}
\thicklines \path(328,1718)(369,1718)
\thicklines \path(2876,1718)(2835,1718)
\put(287,1718){\makebox(0,0)[r]{ 1.98}}
\thicklines \path(328,249)(328,290)
\thicklines \path(328,1718)(328,1677)
\put(328,166){\makebox(0,0){ 2}}
\thicklines \path(611,249)(611,290)
\thicklines \path(611,1718)(611,1677)
\put(611,166){\makebox(0,0){ 4}}
\thicklines \path(894,249)(894,290)
\thicklines \path(894,1718)(894,1677)
\put(894,166){\makebox(0,0){ 6}}
\thicklines \path(1177,249)(1177,290)
\thicklines \path(1177,1718)(1177,1677)
\put(1177,166){\makebox(0,0){ 8}}
\thicklines \path(1460,249)(1460,290)
\thicklines \path(1460,1718)(1460,1677)
\put(1460,166){\makebox(0,0){ 10}}
\thicklines \path(1744,249)(1744,290)
\thicklines \path(1744,1718)(1744,1677)
\put(1744,166){\makebox(0,0){ 12}}
\thicklines \path(2027,249)(2027,290)
\thicklines \path(2027,1718)(2027,1677)
\put(2027,166){\makebox(0,0){ 14}}
\thicklines \path(2310,249)(2310,290)
\thicklines \path(2310,1718)(2310,1677)
\put(2310,166){\makebox(0,0){ 16}}
\thicklines \path(2593,249)(2593,290)
\thicklines \path(2593,1718)(2593,1677)
\put(2593,166){\makebox(0,0){ 18}}
\thicklines \path(2876,249)(2876,290)
\thicklines \path(2876,1718)(2876,1677)
\put(2876,166){\makebox(0,0){ 20}}
\color{black}
\thicklines \path(328,249)(2876,249)(2876,1718)(328,1718)(328,249)
\put(1602,42){\makebox(0,0){$\Re\Delta\omega$}}
\color{red}
\put(470,1141){\makebox(0,0){$\Diamond$}}
\put(611,1218){\makebox(0,0){$\Diamond$}}
\put(753,1280){\makebox(0,0){$\Diamond$}}
\put(894,1326){\makebox(0,0){$\Diamond$}}
\put(1036,1361){\makebox(0,0){$\Diamond$}}
\put(1177,1388){\makebox(0,0){$\Diamond$}}
\put(1319,1409){\makebox(0,0){$\Diamond$}}
\put(1460,1426){\makebox(0,0){$\Diamond$}}
\put(1602,1439){\makebox(0,0){$\Diamond$}}
\put(1744,1450){\makebox(0,0){$\Diamond$}}
\put(1885,1459){\makebox(0,0){$\Diamond$}}
\put(2027,1467){\makebox(0,0){$\Diamond$}}
\put(2168,1473){\makebox(0,0){$\Diamond$}}
\put(2310,1479){\makebox(0,0){$\Diamond$}}
\put(2451,1484){\makebox(0,0){$\Diamond$}}
\put(2593,1488){\makebox(0,0){$\Diamond$}}
\put(2734,1492){\makebox(0,0){$\Diamond$}}
\color{blue}
\thinlines \path(470,1538)(470,1538)(492,1538)(515,1538)(538,1538)(561,1538)(584,1538)(607,1538)(630,1538)(653,1538)(675,1538)(698,1538)(721,1538)(744,1538)(767,1538)(790,1538)(813,1538)(836,1538)(858,1538)(881,1538)(904,1538)(927,1538)(950,1538)(973,1538)(996,1538)(1019,1538)(1041,1538)(1064,1538)(1087,1538)(1110,1538)(1133,1538)(1156,1538)(1179,1538)(1202,1538)(1225,1538)(1247,1538)(1270,1538)(1293,1538)(1316,1538)(1339,1538)(1362,1538)(1385,1538)(1408,1538)(1430,1538)(1453,1538)(1476,1538)(1499,1538)(1522,1538)(1545,1538)(1568,1538)(1591,1538)
\thinlines \path(1591,1538)(1613,1538)(1636,1538)(1659,1538)(1682,1538)(1705,1538)(1728,1538)(1751,1538)(1774,1538)(1796,1538)(1819,1538)(1842,1538)(1865,1538)(1888,1538)(1911,1538)(1934,1538)(1957,1538)(1979,1538)(2002,1538)(2025,1538)(2048,1538)(2071,1538)(2094,1538)(2117,1538)(2140,1538)(2163,1538)(2185,1538)(2208,1538)(2231,1538)(2254,1538)(2277,1538)(2300,1538)(2323,1538)(2346,1538)(2368,1538)(2391,1538)(2414,1538)(2437,1538)(2460,1538)(2483,1538)(2506,1538)(2529,1538)(2551,1538)(2574,1538)(2597,1538)(2620,1538)(2643,1538)(2666,1538)(2689,1538)(2712,1538)(2734,1538)
\color{green}
\thinlines \path(470,375)(470,375)(492,463)(515,540)(538,609)(561,670)(584,725)(607,774)(630,818)(653,859)(675,895)(698,929)(721,959)(744,988)(767,1014)(790,1038)(813,1060)(836,1080)(858,1100)(881,1117)(904,1134)(927,1150)(950,1164)(973,1178)(996,1191)(1019,1203)(1041,1215)(1064,1226)(1087,1236)(1110,1246)(1133,1255)(1156,1263)(1179,1272)(1202,1280)(1225,1287)(1247,1294)(1270,1301)(1293,1308)(1316,1314)(1339,1320)(1362,1325)(1385,1331)(1408,1336)(1430,1341)(1453,1346)(1476,1350)(1499,1355)(1522,1359)(1545,1363)(1568,1367)(1591,1371)
\thinlines \path(1591,1371)(1613,1375)(1636,1378)(1659,1381)(1682,1385)(1705,1388)(1728,1391)(1751,1394)(1774,1397)(1796,1400)(1819,1402)(1842,1405)(1865,1407)(1888,1410)(1911,1412)(1934,1414)(1957,1417)(1979,1419)(2002,1421)(2025,1423)(2048,1425)(2071,1427)(2094,1429)(2117,1430)(2140,1432)(2163,1434)(2185,1436)(2208,1437)(2231,1439)(2254,1440)(2277,1442)(2300,1443)(2323,1445)(2346,1446)(2368,1448)(2391,1449)(2414,1450)(2437,1451)(2460,1453)(2483,1454)(2506,1455)(2529,1456)(2551,1457)(2574,1458)(2597,1459)(2620,1461)(2643,1462)(2666,1463)(2689,1464)(2712,1464)(2734,1465)
\end{picture}
\setlength{\unitlength}{0.0750pt}
\begin{picture}(3000,1800)(0,0)
\footnotesize
\color{black}
\thicklines \path(328,249)(369,249)
\thicklines \path(2876,249)(2835,249)
\put(287,249){\makebox(0,0)[r]{-2.5}}
\thicklines \path(328,433)(369,433)
\thicklines \path(2876,433)(2835,433)
\put(287,433){\makebox(0,0)[r]{-2.48}}
\thicklines \path(328,616)(369,616)
\thicklines \path(2876,616)(2835,616)
\put(287,616){\makebox(0,0)[r]{-2.46}}
\thicklines \path(328,800)(369,800)
\thicklines \path(2876,800)(2835,800)
\put(287,800){\makebox(0,0)[r]{-2.44}}
\thicklines \path(328,984)(369,984)
\thicklines \path(2876,984)(2835,984)
\put(287,984){\makebox(0,0)[r]{-2.42}}
\thicklines \path(328,1167)(369,1167)
\thicklines \path(2876,1167)(2835,1167)
\put(287,1167){\makebox(0,0)[r]{-2.4}}
\thicklines \path(328,1351)(369,1351)
\thicklines \path(2876,1351)(2835,1351)
\put(287,1351){\makebox(0,0)[r]{-2.38}}
\thicklines \path(328,1534)(369,1534)
\thicklines \path(2876,1534)(2835,1534)
\put(287,1534){\makebox(0,0)[r]{-2.36}}
\thicklines \path(328,1718)(369,1718)
\thicklines \path(2876,1718)(2835,1718)
\put(287,1718){\makebox(0,0)[r]{-2.34}}
\thicklines \path(328,249)(328,290)
\thicklines \path(328,1718)(328,1677)
\put(328,166){\makebox(0,0){ 0}}
\thicklines \path(583,249)(583,290)
\thicklines \path(583,1718)(583,1677)
\put(583,166){\makebox(0,0){ 2}}
\thicklines \path(838,249)(838,290)
\thicklines \path(838,1718)(838,1677)
\put(838,166){\makebox(0,0){ 4}}
\thicklines \path(1092,249)(1092,290)
\thicklines \path(1092,1718)(1092,1677)
\put(1092,166){\makebox(0,0){ 6}}
\thicklines \path(1347,249)(1347,290)
\thicklines \path(1347,1718)(1347,1677)
\put(1347,166){\makebox(0,0){ 8}}
\thicklines \path(1602,249)(1602,290)
\thicklines \path(1602,1718)(1602,1677)
\put(1602,166){\makebox(0,0){ 10}}
\thicklines \path(1857,249)(1857,290)
\thicklines \path(1857,1718)(1857,1677)
\put(1857,166){\makebox(0,0){ 12}}
\thicklines \path(2112,249)(2112,290)
\thicklines \path(2112,1718)(2112,1677)
\put(2112,166){\makebox(0,0){ 14}}
\thicklines \path(2366,249)(2366,290)
\thicklines \path(2366,1718)(2366,1677)
\put(2366,166){\makebox(0,0){ 16}}
\thicklines \path(2621,249)(2621,290)
\thicklines \path(2621,1718)(2621,1677)
\put(2621,166){\makebox(0,0){ 18}}
\thicklines \path(2876,249)(2876,290)
\thicklines \path(2876,1718)(2876,1677)
\put(2876,166){\makebox(0,0){ 20}}
\color{black}
\thicklines \path(328,249)(2876,249)(2876,1718)(328,1718)(328,249)
\put(1602,42){\makebox(0,0){$\Im\Delta\omega$}}
\color{red}
\put(583,304){\makebox(0,0){$\Diamond$}}
\put(710,1245){\makebox(0,0){$\Diamond$}}
\put(838,1436){\makebox(0,0){$\Diamond$}}
\put(965,1507){\makebox(0,0){$\Diamond$}}
\put(1092,1542){\makebox(0,0){$\Diamond$}}
\put(1220,1562){\makebox(0,0){$\Diamond$}}
\put(1347,1575){\makebox(0,0){$\Diamond$}}
\put(1475,1584){\makebox(0,0){$\Diamond$}}
\put(1602,1590){\makebox(0,0){$\Diamond$}}
\put(1729,1595){\makebox(0,0){$\Diamond$}}
\put(1857,1599){\makebox(0,0){$\Diamond$}}
\put(1984,1602){\makebox(0,0){$\Diamond$}}
\put(2112,1604){\makebox(0,0){$\Diamond$}}
\put(2239,1607){\makebox(0,0){$\Diamond$}}
\put(2366,1608){\makebox(0,0){$\Diamond$}}
\put(2494,1610){\makebox(0,0){$\Diamond$}}
\put(2621,1611){\makebox(0,0){$\Diamond$}}
\put(2749,1612){\makebox(0,0){$\Diamond$}}
\color{blue}
\thinlines \path(583,1626)(583,1626)(605,1626)(627,1626)(648,1626)(670,1626)(692,1626)(714,1626)(736,1626)(758,1626)(780,1626)(802,1626)(823,1626)(845,1626)(867,1626)(889,1626)(911,1626)(933,1626)(955,1626)(977,1626)(998,1626)(1020,1626)(1042,1626)(1064,1626)(1086,1626)(1108,1626)(1130,1626)(1152,1626)(1173,1626)(1195,1626)(1217,1626)(1239,1626)(1261,1626)(1283,1626)(1305,1626)(1327,1626)(1348,1626)(1370,1626)(1392,1626)(1414,1626)(1436,1626)(1458,1626)(1480,1626)(1502,1626)(1524,1626)(1545,1626)(1567,1626)(1589,1626)(1611,1626)(1633,1626)(1655,1626)
\thinlines \path(1655,1626)(1677,1626)(1699,1626)(1720,1626)(1742,1626)(1764,1626)(1786,1626)(1808,1626)(1830,1626)(1852,1626)(1874,1626)(1895,1626)(1917,1626)(1939,1626)(1961,1626)(1983,1626)(2005,1626)(2027,1626)(2049,1626)(2070,1626)(2092,1626)(2114,1626)(2136,1626)(2158,1626)(2180,1626)(2202,1626)(2224,1626)(2245,1626)(2267,1626)(2289,1626)(2311,1626)(2333,1626)(2355,1626)(2377,1626)(2399,1626)(2420,1626)(2442,1626)(2464,1626)(2486,1626)(2508,1626)(2530,1626)(2552,1626)(2574,1626)(2595,1626)(2617,1626)(2639,1626)(2661,1626)(2683,1626)(2705,1626)(2727,1626)(2749,1626)
\color{green}
\thinlines \path(583,784)(583,784)(605,882)(627,962)(648,1029)(670,1085)(692,1133)(714,1175)(736,1210)(758,1242)(780,1269)(802,1294)(823,1316)(845,1335)(867,1353)(889,1368)(911,1383)(933,1396)(955,1408)(977,1419)(998,1429)(1020,1438)(1042,1447)(1064,1455)(1086,1462)(1108,1469)(1130,1475)(1152,1481)(1173,1487)(1195,1492)(1217,1497)(1239,1502)(1261,1506)(1283,1510)(1305,1514)(1327,1518)(1348,1521)(1370,1524)(1392,1528)(1414,1530)(1436,1533)(1458,1536)(1480,1539)(1502,1541)(1524,1543)(1545,1546)(1567,1548)(1589,1550)(1611,1552)(1633,1554)(1655,1555)
\thinlines \path(1655,1555)(1677,1557)(1699,1559)(1720,1560)(1742,1562)(1764,1563)(1786,1565)(1808,1566)(1830,1567)(1852,1569)(1874,1570)(1895,1571)(1917,1572)(1939,1573)(1961,1574)(1983,1575)(2005,1576)(2027,1577)(2049,1578)(2070,1579)(2092,1580)(2114,1581)(2136,1582)(2158,1582)(2180,1583)(2202,1584)(2224,1585)(2245,1585)(2267,1586)(2289,1587)(2311,1587)(2333,1588)(2355,1589)(2377,1589)(2399,1590)(2420,1590)(2442,1591)(2464,1591)(2486,1592)(2508,1593)(2530,1593)(2552,1594)(2574,1594)(2595,1594)(2617,1595)(2639,1595)(2661,1596)(2683,1596)(2705,1597)(2727,1597)(2749,1597)
\end{picture}
\caption{\label{fig4}The frequency gap~(\ref{eqgap}) for vector perturbations in $d=4$ for $r_H=1$ and $\ell = 2$: zeroth and first order analytical (eq.~(\ref{eq118})) compared with numerical data~\cite{CKL}.}
\end{center}
\end{figure}
Numerically for large $r_H$ and $\ell =2$,
\be \frac{\o_n}{r_H} = (1.299-2.250i)n +0.573-0.419i +\frac{8.19+6.29i}{r_H^2 \sqrt n} \ee
For an intermediate black hole, $r_H = 1$, we obtain (see figure~\ref{fig4})
\be\label{eq118} \o_n = (1.969-2.350i)n + 0.752-0.370i +\frac{0.741+0.519i}{\sqrt n} \ee
and for a small black hole, $r_H = 0.2$, we obtain (see figure~\ref{fig5})
\be\label{eq119} \o_n = (1.695-0.571i)n + 0.487-0.0441i +\frac{1.239+0.6357i}{\sqrt n} \ee
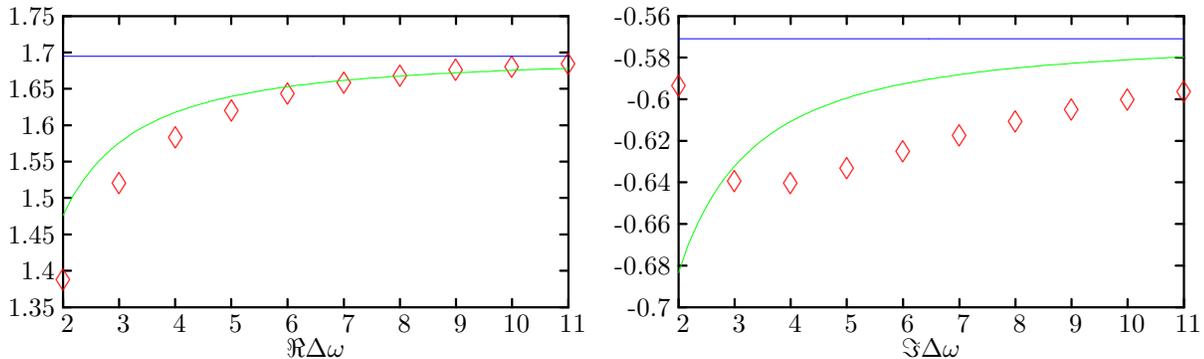
\begin{figure}
\begin{center}
\setlength{\unitlength}{0.0750pt}
\begin{picture}(3000,1800)(0,0)
\footnotesize
\color{black}
\thicklines \path(328,249)(369,249)
\thicklines \path(2876,249)(2835,249)
\put(287,249){\makebox(0,0)[r]{ 1.35}}
\thicklines \path(328,433)(369,433)
\thicklines \path(2876,433)(2835,433)
\put(287,433){\makebox(0,0)[r]{ 1.4}}
\thicklines \path(328,616)(369,616)
\thicklines \path(2876,616)(2835,616)
\put(287,616){\makebox(0,0)[r]{ 1.45}}
\thicklines \path(328,800)(369,800)
\thicklines \path(2876,800)(2835,800)
\put(287,800){\makebox(0,0)[r]{ 1.5}}
\thicklines \path(328,984)(369,984)
\thicklines \path(2876,984)(2835,984)
\put(287,984){\makebox(0,0)[r]{ 1.55}}
\thicklines \path(328,1167)(369,1167)
\thicklines \path(2876,1167)(2835,1167)
\put(287,1167){\makebox(0,0)[r]{ 1.6}}
\thicklines \path(328,1351)(369,1351)
\thicklines \path(2876,1351)(2835,1351)
\put(287,1351){\makebox(0,0)[r]{ 1.65}}
\thicklines \path(328,1534)(369,1534)
\thicklines \path(2876,1534)(2835,1534)
\put(287,1534){\makebox(0,0)[r]{ 1.7}}
\thicklines \path(328,1718)(369,1718)
\thicklines \path(2876,1718)(2835,1718)
\put(287,1718){\makebox(0,0)[r]{ 1.75}}
\thicklines \path(328,249)(328,290)
\thicklines \path(328,1718)(328,1677)
\put(328,166){\makebox(0,0){ 2}}
\thicklines \path(611,249)(611,290)
\thicklines \path(611,1718)(611,1677)
\put(611,166){\makebox(0,0){ 3}}
\thicklines \path(894,249)(894,290)
\thicklines \path(894,1718)(894,1677)
\put(894,166){\makebox(0,0){ 4}}
\thicklines \path(1177,249)(1177,290)
\thicklines \path(1177,1718)(1177,1677)
\put(1177,166){\makebox(0,0){ 5}}
\thicklines \path(1460,249)(1460,290)
\thicklines \path(1460,1718)(1460,1677)
\put(1460,166){\makebox(0,0){ 6}}
\thicklines \path(1744,249)(1744,290)
\thicklines \path(1744,1718)(1744,1677)
\put(1744,166){\makebox(0,0){ 7}}
\thicklines \path(2027,249)(2027,290)
\thicklines \path(2027,1718)(2027,1677)
\put(2027,166){\makebox(0,0){ 8}}
\thicklines \path(2310,249)(2310,290)
\thicklines \path(2310,1718)(2310,1677)
\put(2310,166){\makebox(0,0){ 9}}
\thicklines \path(2593,249)(2593,290)
\thicklines \path(2593,1718)(2593,1677)
\put(2593,166){\makebox(0,0){ 10}}
\thicklines \path(2876,249)(2876,290)
\thicklines \path(2876,1718)(2876,1677)
\put(2876,166){\makebox(0,0){ 11}}
\color{black}
\thicklines \path(328,249)(2876,249)(2876,1718)(328,1718)(328,249)
\put(1602,42){\makebox(0,0){$\Re\Delta\omega$}}
\color{red}
\put(328,386){\makebox(0,0){$\Diamond$}}
\put(611,873){\makebox(0,0){$\Diamond$}}
\put(894,1108){\makebox(0,0){$\Diamond$}}
\put(1177,1243){\makebox(0,0){$\Diamond$}}
\put(1460,1328){\makebox(0,0){$\Diamond$}}
\put(1744,1382){\makebox(0,0){$\Diamond$}}
\put(2027,1419){\makebox(0,0){$\Diamond$}}
\put(2310,1444){\makebox(0,0){$\Diamond$}}
\put(2593,1462){\makebox(0,0){$\Diamond$}}
\put(2876,1475){\makebox(0,0){$\Diamond$}}
\color{blue}
\thinlines \path(328,1516)(328,1516)(354,1516)(379,1516)(405,1516)(431,1516)(457,1516)(482,1516)(508,1516)(534,1516)(560,1516)(585,1516)(611,1516)(637,1516)(663,1516)(688,1516)(714,1516)(740,1516)(766,1516)(791,1516)(817,1516)(843,1516)(868,1516)(894,1516)(920,1516)(946,1516)(971,1516)(997,1516)(1023,1516)(1049,1516)(1074,1516)(1100,1516)(1126,1516)(1152,1516)(1177,1516)(1203,1516)(1229,1516)(1255,1516)(1280,1516)(1306,1516)(1332,1516)(1357,1516)(1383,1516)(1409,1516)(1435,1516)(1460,1516)(1486,1516)(1512,1516)(1538,1516)(1563,1516)(1589,1516)
\thinlines \path(1589,1516)(1615,1516)(1641,1516)(1666,1516)(1692,1516)(1718,1516)(1744,1516)(1769,1516)(1795,1516)(1821,1516)(1847,1516)(1872,1516)(1898,1516)(1924,1516)(1949,1516)(1975,1516)(2001,1516)(2027,1516)(2052,1516)(2078,1516)(2104,1516)(2130,1516)(2155,1516)(2181,1516)(2207,1516)(2233,1516)(2258,1516)(2284,1516)(2310,1516)(2336,1516)(2361,1516)(2387,1516)(2413,1516)(2438,1516)(2464,1516)(2490,1516)(2516,1516)(2541,1516)(2567,1516)(2593,1516)(2619,1516)(2644,1516)(2670,1516)(2696,1516)(2722,1516)(2747,1516)(2773,1516)(2799,1516)(2825,1516)(2850,1516)(2876,1516)
\color{green}
\thinlines \path(328,712)(328,712)(354,764)(379,810)(405,852)(431,890)(457,924)(482,956)(508,985)(534,1011)(560,1035)(585,1057)(611,1078)(637,1097)(663,1115)(688,1132)(714,1147)(740,1162)(766,1175)(791,1188)(817,1200)(843,1211)(868,1222)(894,1232)(920,1241)(946,1250)(971,1258)(997,1266)(1023,1274)(1049,1281)(1074,1288)(1100,1295)(1126,1301)(1152,1307)(1177,1313)(1203,1318)(1229,1323)(1255,1328)(1280,1333)(1306,1337)(1332,1342)(1357,1346)(1383,1350)(1409,1354)(1435,1358)(1460,1361)(1486,1365)(1512,1368)(1538,1371)(1563,1374)(1589,1377)
\thinlines \path(1589,1377)(1615,1380)(1641,1383)(1666,1386)(1692,1388)(1718,1391)(1744,1393)(1769,1396)(1795,1398)(1821,1400)(1847,1402)(1872,1404)(1898,1406)(1924,1408)(1949,1410)(1975,1412)(2001,1414)(2027,1415)(2052,1417)(2078,1419)(2104,1420)(2130,1422)(2155,1423)(2181,1425)(2207,1426)(2233,1428)(2258,1429)(2284,1430)(2310,1432)(2336,1433)(2361,1434)(2387,1435)(2413,1437)(2438,1438)(2464,1439)(2490,1440)(2516,1441)(2541,1442)(2567,1443)(2593,1444)(2619,1445)(2644,1446)(2670,1447)(2696,1448)(2722,1449)(2747,1450)(2773,1450)(2799,1451)(2825,1452)(2850,1453)(2876,1454)
\end{picture}
\setlength{\unitlength}{0.0750pt}
\begin{picture}(3000,1800)(0,0)
\footnotesize
\color{black}
\thicklines \path(328,249)(369,249)
\thicklines \path(2876,249)(2835,249)
\put(287,249){\makebox(0,0)[r]{-0.7}}
\thicklines \path(328,459)(369,459)
\thicklines \path(2876,459)(2835,459)
\put(287,459){\makebox(0,0)[r]{-0.68}}
\thicklines \path(328,669)(369,669)
\thicklines \path(2876,669)(2835,669)
\put(287,669){\makebox(0,0)[r]{-0.66}}
\thicklines \path(328,879)(369,879)
\thicklines \path(2876,879)(2835,879)
\put(287,879){\makebox(0,0)[r]{-0.64}}
\thicklines \path(328,1088)(369,1088)
\thicklines \path(2876,1088)(2835,1088)
\put(287,1088){\makebox(0,0)[r]{-0.62}}
\thicklines \path(328,1298)(369,1298)
\thicklines \path(2876,1298)(2835,1298)
\put(287,1298){\makebox(0,0)[r]{-0.6}}
\thicklines \path(328,1508)(369,1508)
\thicklines \path(2876,1508)(2835,1508)
\put(287,1508){\makebox(0,0)[r]{-0.58}}
\thicklines \path(328,1718)(369,1718)
\thicklines \path(2876,1718)(2835,1718)
\put(287,1718){\makebox(0,0)[r]{-0.56}}
\thicklines \path(328,249)(328,290)
\thicklines \path(328,1718)(328,1677)
\put(328,166){\makebox(0,0){ 2}}
\thicklines \path(611,249)(611,290)
\thicklines \path(611,1718)(611,1677)
\put(611,166){\makebox(0,0){ 3}}
\thicklines \path(894,249)(894,290)
\thicklines \path(894,1718)(894,1677)
\put(894,166){\makebox(0,0){ 4}}
\thicklines \path(1177,249)(1177,290)
\thicklines \path(1177,1718)(1177,1677)
\put(1177,166){\makebox(0,0){ 5}}
\thicklines \path(1460,249)(1460,290)
\thicklines \path(1460,1718)(1460,1677)
\put(1460,166){\makebox(0,0){ 6}}
\thicklines \path(1744,249)(1744,290)
\thicklines \path(1744,1718)(1744,1677)
\put(1744,166){\makebox(0,0){ 7}}
\thicklines \path(2027,249)(2027,290)
\thicklines \path(2027,1718)(2027,1677)
\put(2027,166){\makebox(0,0){ 8}}
\thicklines \path(2310,249)(2310,290)
\thicklines \path(2310,1718)(2310,1677)
\put(2310,166){\makebox(0,0){ 9}}
\thicklines \path(2593,249)(2593,290)
\thicklines \path(2593,1718)(2593,1677)
\put(2593,166){\makebox(0,0){ 10}}
\thicklines \path(2876,249)(2876,290)
\thicklines \path(2876,1718)(2876,1677)
\put(2876,166){\makebox(0,0){ 11}}
\color{black}
\thicklines \path(328,249)(2876,249)(2876,1718)(328,1718)(328,249)
\put(1602,42){\makebox(0,0){$\Im\Delta\omega$}}
\color{red}
\put(328,1365){\makebox(0,0){$\Diamond$}}
\put(611,884){\makebox(0,0){$\Diamond$}}
\put(894,877){\makebox(0,0){$\Diamond$}}
\put(1177,948){\makebox(0,0){$\Diamond$}}
\put(1460,1033){\makebox(0,0){$\Diamond$}}
\put(1744,1114){\makebox(0,0){$\Diamond$}}
\put(2027,1185){\makebox(0,0){$\Diamond$}}
\put(2310,1245){\makebox(0,0){$\Diamond$}}
\put(2593,1296){\makebox(0,0){$\Diamond$}}
\put(2876,1338){\makebox(0,0){$\Diamond$}}
\color{blue}
\thinlines \path(328,1603)(328,1603)(354,1603)(379,1603)(405,1603)(431,1603)(457,1603)(482,1603)(508,1603)(534,1603)(560,1603)(585,1603)(611,1603)(637,1603)(663,1603)(688,1603)(714,1603)(740,1603)(766,1603)(791,1603)(817,1603)(843,1603)(868,1603)(894,1603)(920,1603)(946,1603)(971,1603)(997,1603)(1023,1603)(1049,1603)(1074,1603)(1100,1603)(1126,1603)(1152,1603)(1177,1603)(1203,1603)(1229,1603)(1255,1603)(1280,1603)(1306,1603)(1332,1603)(1357,1603)(1383,1603)(1409,1603)(1435,1603)(1460,1603)(1486,1603)(1512,1603)(1538,1603)(1563,1603)(1589,1603)
\thinlines \path(1589,1603)(1615,1603)(1641,1603)(1666,1603)(1692,1603)(1718,1603)(1744,1603)(1769,1603)(1795,1603)(1821,1603)(1847,1603)(1872,1603)(1898,1603)(1924,1603)(1949,1603)(1975,1603)(2001,1603)(2027,1603)(2052,1603)(2078,1603)(2104,1603)(2130,1603)(2155,1603)(2181,1603)(2207,1603)(2233,1603)(2258,1603)(2284,1603)(2310,1603)(2336,1603)(2361,1603)(2387,1603)(2413,1603)(2438,1603)(2464,1603)(2490,1603)(2516,1603)(2541,1603)(2567,1603)(2593,1603)(2619,1603)(2644,1603)(2670,1603)(2696,1603)(2722,1603)(2747,1603)(2773,1603)(2799,1603)(2825,1603)(2850,1603)(2876,1603)
\color{green}
\thinlines \path(328,423)(328,423)(354,499)(379,568)(405,629)(431,685)(457,735)(482,781)(508,823)(534,862)(560,898)(585,930)(611,961)(637,989)(663,1015)(688,1039)(714,1062)(740,1083)(766,1103)(791,1122)(817,1139)(843,1156)(868,1171)(894,1186)(920,1200)(946,1213)(971,1225)(997,1237)(1023,1248)(1049,1258)(1074,1268)(1100,1278)(1126,1287)(1152,1296)(1177,1304)(1203,1312)(1229,1320)(1255,1327)(1280,1334)(1306,1341)(1332,1347)(1357,1353)(1383,1359)(1409,1365)(1435,1370)(1460,1376)(1486,1381)(1512,1386)(1538,1390)(1563,1395)(1589,1399)
\thinlines \path(1589,1399)(1615,1403)(1641,1407)(1666,1411)(1692,1415)(1718,1419)(1744,1422)(1769,1426)(1795,1429)(1821,1433)(1847,1436)(1872,1439)(1898,1442)(1924,1445)(1949,1447)(1975,1450)(2001,1453)(2027,1455)(2052,1458)(2078,1460)(2104,1462)(2130,1465)(2155,1467)(2181,1469)(2207,1471)(2233,1473)(2258,1475)(2284,1477)(2310,1479)(2336,1481)(2361,1483)(2387,1484)(2413,1486)(2438,1488)(2464,1489)(2490,1491)(2516,1493)(2541,1494)(2567,1496)(2593,1497)(2619,1499)(2644,1500)(2670,1501)(2696,1503)(2722,1504)(2747,1505)(2773,1506)(2799,1508)(2825,1509)(2850,1510)(2876,1511)
\end{picture}
\caption{\label{fig5}The frequency gap~(\ref{eqgap}) for vector perturbations in $d=4$ for $r_H=0.2$ and $\ell = 2$: zeroth and first order analytical (eq.~(\ref{eq119})) compared with numerical data~\cite{CKL}.}
\end{center}
\end{figure}
In all cases of gravitational perturbations, regardless of the size of the
black hole, our analytical results are in good agreement with numerical results~\cite{CKL}.

\section{Electromagnetic perturbations}\label{sect3}
In this section we extend the discussion to electromagnetic perturbations.
This is a singular case because the potential vanishes at zeroth order.
Consequently, the compatibility condition~(\ref{eqcomp}) discussed in the
previous section has no solutions and no asymptotic expression for quasi-normal
frequencies may be deduced~\cite{NS}.
Nevertheless, the numerical results are similar to the ones we discussed in
the case of gravitational perturbations~\cite{CKL}.
We shall show that including first-order corrections leads to analytical asymptotic
expressions for quasi-normal frequencies in agreement with numerical results.
Unlike with gravitational perturbations, where first-order corrections were a
power of $n$ (eqs.~(\ref{eqo1st}) and (\ref{eqab})), for electromagnetic perturbations first-order
corrections are $o(\ln n)$.

We shall concentrate on the four-dimensional case for definiteness.
Generalization to higher dimensions is straightforward.
The wave equation reduces to~(\ref{sch}) with electromagnetic potential
\be\label{V-V}
  V_{\mathsf{EM}} =\frac{\ell(\ell+1)}{r^2}f(r).
\ee
where $f(r)$ is given in~(\ref{line}) with $d=4$.
Near the origin, this potential may be expanded in terms of the tortoise coordinate. Using eq.~(\ref{x0}), we obtain
\be\label{eqVEMr}
  V_{\mathsf{EM}} =\frac{j^2-1}{4r_*^2}+\frac{\ell(\ell+1)r_*^{-3/2}}{2\sqrt{-4\mu}}+\dots \ ,
\ee
where $j=1$. This leads to a vanishing potential to zeroth order.
Consequently, no analytic expression for quasi-normal frequencies is deduced.
This is easily seen by substituting $j=1$ in the zeroth-order expression~(\ref{omega-0});
we obtain a divergent result because $\tan^{-1} i$ is not finite.

This is remedied by including first-order corrections.
The compatibility condition~(\ref{eq99}) of the two first-order constraints (\ref{newconstr1a}) and
(\ref{newconstr2}) reads
\be
\left|
\begin{array}{cc}
1+\bar b+\bar a_2\tan\omega \bar r_* & -i-\bar b+i\bar a_2 \\
(1-\bar b)\tan\omega \bar r_* -\bar a_1 & 1-\bar a_1+i\bar b 
\end{array}
\right| = 0
\ee
where we used $d=4$, $j=1$, $j_\infty = d-3=1$, and $\alpha_+=\beta = \frac{\pi}{2}$.
At zeroth order (setting $\bar a_1 = \bar a_2 = \bar b = 0$), we obtain
\be \tan \o \bar r_* = i \ee
which has no solution, as expected~\cite{NS}.
At first order, we obtain
\be \tan \o \bar r_* = i+(1-i)(\bar a_1 - \bar a_2
- 2\bar b)\ee
whose first-order solution is
\be\label{eqEMo}
\omega {\bar r}_* = n\pi+\frac{1}{2i}\ln \frac{(1+i)(\bar a_1 - \bar a_2
- 2\bar b)}{2}
\ee
Using~(\ref{eqab}) and (\ref{eqVEMr}), we deduce explicit expressions for the
first-order coefficients,
\be \bar a_1 = \mathcal{A} \sqrt{\frac{\bar r_*}{n}} \ , \ \ \ \ \bar a_2 = \bar b = -\bar a_1  \ , \ \ \ \ \mathcal{A}
= \frac{\ell(\ell+1)}{2\sqrt{-4\mu}} \ee
and eq.~(\ref{eqEMo}) reads explicitly
\be\label{eqEMo1}
\omega {\bar r}_* = n\pi -\frac{i}{4}\ln n+\frac{1}{2i}\ln\left( 2(1+i){\cal A}\sqrt{\bar r_*}\right)
\ee
Therefore, the correction to the quasi-normal frequencies behaves as $\ln n$ in the large $\o$ limit.

To compare with numerical results, set $R=1$.
As with gravitational perturbations, we shall compare the gap, because the
offset is not reliable.
For the gap, we have from~(\ref{eqEMo1})
\be \Delta\o_n \equiv \omega_n - \omega_{n-1} = \frac{\pi}{\bar r_*} \left( 1- \frac{i}{4\pi n} + \dots \right) \ee
Both leading and sub-leading terms are independent of $\ell$.

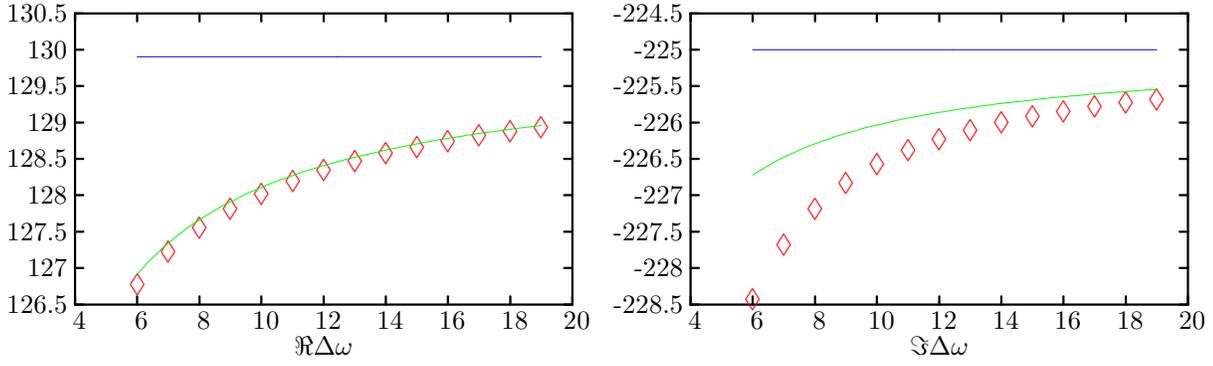
\begin{figure}
\begin{center}
\setlength{\unitlength}{0.0750pt}
\begin{picture}(3000,1800)(0,0)
\footnotesize
\color{black}
\thicklines \path(369,249)(410,249)
\thicklines \path(2876,249)(2835,249)
\put(328,249){\makebox(0,0)[r]{ 126.5}}
\thicklines \path(369,433)(410,433)
\thicklines \path(2876,433)(2835,433)
\put(328,433){\makebox(0,0)[r]{ 127}}
\thicklines \path(369,616)(410,616)
\thicklines \path(2876,616)(2835,616)
\put(328,616){\makebox(0,0)[r]{ 127.5}}
\thicklines \path(369,800)(410,800)
\thicklines \path(2876,800)(2835,800)
\put(328,800){\makebox(0,0)[r]{ 128}}
\thicklines \path(369,984)(410,984)
\thicklines \path(2876,984)(2835,984)
\put(328,984){\makebox(0,0)[r]{ 128.5}}
\thicklines \path(369,1167)(410,1167)
\thicklines \path(2876,1167)(2835,1167)
\put(328,1167){\makebox(0,0)[r]{ 129}}
\thicklines \path(369,1351)(410,1351)
\thicklines \path(2876,1351)(2835,1351)
\put(328,1351){\makebox(0,0)[r]{ 129.5}}
\thicklines \path(369,1534)(410,1534)
\thicklines \path(2876,1534)(2835,1534)
\put(328,1534){\makebox(0,0)[r]{ 130}}
\thicklines \path(369,1718)(410,1718)
\thicklines \path(2876,1718)(2835,1718)
\put(328,1718){\makebox(0,0)[r]{ 130.5}}
\thicklines \path(369,249)(369,290)
\thicklines \path(369,1718)(369,1677)
\put(369,166){\makebox(0,0){ 4}}
\thicklines \path(682,249)(682,290)
\thicklines \path(682,1718)(682,1677)
\put(682,166){\makebox(0,0){ 6}}
\thicklines \path(996,249)(996,290)
\thicklines \path(996,1718)(996,1677)
\put(996,166){\makebox(0,0){ 8}}
\thicklines \path(1309,249)(1309,290)
\thicklines \path(1309,1718)(1309,1677)
\put(1309,166){\makebox(0,0){ 10}}
\thicklines \path(1623,249)(1623,290)
\thicklines \path(1623,1718)(1623,1677)
\put(1623,166){\makebox(0,0){ 12}}
\thicklines \path(1936,249)(1936,290)
\thicklines \path(1936,1718)(1936,1677)
\put(1936,166){\makebox(0,0){ 14}}
\thicklines \path(2249,249)(2249,290)
\thicklines \path(2249,1718)(2249,1677)
\put(2249,166){\makebox(0,0){ 16}}
\thicklines \path(2563,249)(2563,290)
\thicklines \path(2563,1718)(2563,1677)
\put(2563,166){\makebox(0,0){ 18}}
\thicklines \path(2876,249)(2876,290)
\thicklines \path(2876,1718)(2876,1677)
\put(2876,166){\makebox(0,0){ 20}}
\color{black}
\thicklines \path(369,249)(2876,249)(2876,1718)(369,1718)(369,249)
\put(1622,42){\makebox(0,0){$\Re\Delta\omega$}}
\color{red}
\put(682,349){\makebox(0,0){$\Diamond$}}
\put(839,513){\makebox(0,0){$\Diamond$}}
\put(996,633){\makebox(0,0){$\Diamond$}}
\put(1152,729){\makebox(0,0){$\Diamond$}}
\put(1309,807){\makebox(0,0){$\Diamond$}}
\put(1466,871){\makebox(0,0){$\Diamond$}}
\put(1623,925){\makebox(0,0){$\Diamond$}}
\put(1779,971){\makebox(0,0){$\Diamond$}}
\put(1936,1010){\makebox(0,0){$\Diamond$}}
\put(2093,1044){\makebox(0,0){$\Diamond$}}
\put(2249,1073){\makebox(0,0){$\Diamond$}}
\put(2406,1099){\makebox(0,0){$\Diamond$}}
\put(2563,1123){\makebox(0,0){$\Diamond$}}
\put(2719,1143){\makebox(0,0){$\Diamond$}}
\color{blue}
\thinlines \path(682,1498)(682,1498)(703,1498)(724,1498)(744,1498)(765,1498)(785,1498)(806,1498)(826,1498)(847,1498)(868,1498)(888,1498)(909,1498)(929,1498)(950,1498)(970,1498)(991,1498)(1012,1498)(1032,1498)(1053,1498)(1073,1498)(1094,1498)(1114,1498)(1135,1498)(1156,1498)(1176,1498)(1197,1498)(1217,1498)(1238,1498)(1258,1498)(1279,1498)(1300,1498)(1320,1498)(1341,1498)(1361,1498)(1382,1498)(1403,1498)(1423,1498)(1444,1498)(1464,1498)(1485,1498)(1505,1498)(1526,1498)(1547,1498)(1567,1498)(1588,1498)(1608,1498)(1629,1498)(1649,1498)(1670,1498)(1691,1498)
\thinlines \path(1691,1498)(1711,1498)(1732,1498)(1752,1498)(1773,1498)(1793,1498)(1814,1498)(1835,1498)(1855,1498)(1876,1498)(1896,1498)(1917,1498)(1937,1498)(1958,1498)(1979,1498)(1999,1498)(2020,1498)(2040,1498)(2061,1498)(2081,1498)(2102,1498)(2123,1498)(2143,1498)(2164,1498)(2184,1498)(2205,1498)(2226,1498)(2246,1498)(2267,1498)(2287,1498)(2308,1498)(2328,1498)(2349,1498)(2370,1498)(2390,1498)(2411,1498)(2431,1498)(2452,1498)(2472,1498)(2493,1498)(2514,1498)(2534,1498)(2555,1498)(2575,1498)(2596,1498)(2616,1498)(2637,1498)(2658,1498)(2678,1498)(2699,1498)(2719,1498)
\color{green}
\thinlines \path(682,402)(682,402)(703,425)(724,448)(744,470)(765,490)(785,510)(806,529)(826,548)(847,565)(868,582)(888,599)(909,615)(929,630)(950,645)(970,659)(991,673)(1012,686)(1032,699)(1053,712)(1073,724)(1094,736)(1114,747)(1135,758)(1156,769)(1176,779)(1197,789)(1217,799)(1238,809)(1258,818)(1279,827)(1300,836)(1320,845)(1341,853)(1361,861)(1382,869)(1403,877)(1423,885)(1444,892)(1464,899)(1485,907)(1505,913)(1526,920)(1547,927)(1567,933)(1588,939)(1608,946)(1629,952)(1649,958)(1670,963)(1691,969)
\thinlines \path(1691,969)(1711,974)(1732,980)(1752,985)(1773,990)(1793,995)(1814,1000)(1835,1005)(1855,1010)(1876,1015)(1896,1019)(1917,1024)(1937,1028)(1958,1033)(1979,1037)(1999,1041)(2020,1045)(2040,1049)(2061,1053)(2081,1057)(2102,1061)(2123,1065)(2143,1069)(2164,1072)(2184,1076)(2205,1079)(2226,1083)(2246,1086)(2267,1090)(2287,1093)(2308,1096)(2328,1099)(2349,1103)(2370,1106)(2390,1109)(2411,1112)(2431,1115)(2452,1118)(2472,1120)(2493,1123)(2514,1126)(2534,1129)(2555,1131)(2575,1134)(2596,1137)(2616,1139)(2637,1142)(2658,1144)(2678,1147)(2699,1149)(2719,1152)
\end{picture}
\setlength{\unitlength}{0.0750pt}
\begin{picture}(3000,1800)(0,0)
\footnotesize
\color{black}
\thicklines \path(369,249)(410,249)
\thicklines \path(2876,249)(2835,249)
\put(328,249){\makebox(0,0)[r]{-228.5}}
\thicklines \path(369,433)(410,433)
\thicklines \path(2876,433)(2835,433)
\put(328,433){\makebox(0,0)[r]{-228}}
\thicklines \path(369,616)(410,616)
\thicklines \path(2876,616)(2835,616)
\put(328,616){\makebox(0,0)[r]{-227.5}}
\thicklines \path(369,800)(410,800)
\thicklines \path(2876,800)(2835,800)
\put(328,800){\makebox(0,0)[r]{-227}}
\thicklines \path(369,984)(410,984)
\thicklines \path(2876,984)(2835,984)
\put(328,984){\makebox(0,0)[r]{-226.5}}
\thicklines \path(369,1167)(410,1167)
\thicklines \path(2876,1167)(2835,1167)
\put(328,1167){\makebox(0,0)[r]{-226}}
\thicklines \path(369,1351)(410,1351)
\thicklines \path(2876,1351)(2835,1351)
\put(328,1351){\makebox(0,0)[r]{-225.5}}
\thicklines \path(369,1534)(410,1534)
\thicklines \path(2876,1534)(2835,1534)
\put(328,1534){\makebox(0,0)[r]{-225}}
\thicklines \path(369,1718)(410,1718)
\thicklines \path(2876,1718)(2835,1718)
\put(328,1718){\makebox(0,0)[r]{-224.5}}
\thicklines \path(369,249)(369,290)
\thicklines \path(369,1718)(369,1677)
\put(369,166){\makebox(0,0){ 4}}
\thicklines \path(682,249)(682,290)
\thicklines \path(682,1718)(682,1677)
\put(682,166){\makebox(0,0){ 6}}
\thicklines \path(996,249)(996,290)
\thicklines \path(996,1718)(996,1677)
\put(996,166){\makebox(0,0){ 8}}
\thicklines \path(1309,249)(1309,290)
\thicklines \path(1309,1718)(1309,1677)
\put(1309,166){\makebox(0,0){ 10}}
\thicklines \path(1623,249)(1623,290)
\thicklines \path(1623,1718)(1623,1677)
\put(1623,166){\makebox(0,0){ 12}}
\thicklines \path(1936,249)(1936,290)
\thicklines \path(1936,1718)(1936,1677)
\put(1936,166){\makebox(0,0){ 14}}
\thicklines \path(2249,249)(2249,290)
\thicklines \path(2249,1718)(2249,1677)
\put(2249,166){\makebox(0,0){ 16}}
\thicklines \path(2563,249)(2563,290)
\thicklines \path(2563,1718)(2563,1677)
\put(2563,166){\makebox(0,0){ 18}}
\thicklines \path(2876,249)(2876,290)
\thicklines \path(2876,1718)(2876,1677)
\put(2876,166){\makebox(0,0){ 20}}
\color{black}
\thicklines \path(369,249)(2876,249)(2876,1718)(369,1718)(369,249)
\put(1622,42){\makebox(0,0){$\Im\Delta\omega$}}
\color{red}
\put(682,271){\makebox(0,0){$\Diamond$}}
\put(839,548){\makebox(0,0){$\Diamond$}}
\put(996,732){\makebox(0,0){$\Diamond$}}
\put(1152,860){\makebox(0,0){$\Diamond$}}
\put(1309,954){\makebox(0,0){$\Diamond$}}
\put(1466,1027){\makebox(0,0){$\Diamond$}}
\put(1623,1083){\makebox(0,0){$\Diamond$}}
\put(1779,1129){\makebox(0,0){$\Diamond$}}
\put(1936,1166){\makebox(0,0){$\Diamond$}}
\put(2093,1197){\makebox(0,0){$\Diamond$}}
\put(2249,1224){\makebox(0,0){$\Diamond$}}
\put(2406,1246){\makebox(0,0){$\Diamond$}}
\put(2563,1266){\makebox(0,0){$\Diamond$}}
\put(2719,1283){\makebox(0,0){$\Diamond$}}
\color{blue}
\thinlines \path(682,1534)(682,1534)(703,1534)(724,1534)(744,1534)(765,1534)(785,1534)(806,1534)(826,1534)(847,1534)(868,1534)(888,1534)(909,1534)(929,1534)(950,1534)(970,1534)(991,1534)(1012,1534)(1032,1534)(1053,1534)(1073,1534)(1094,1534)(1114,1534)(1135,1534)(1156,1534)(1176,1534)(1197,1534)(1217,1534)(1238,1534)(1258,1534)(1279,1534)(1300,1534)(1320,1534)(1341,1534)(1361,1534)(1382,1534)(1403,1534)(1423,1534)(1444,1534)(1464,1534)(1485,1534)(1505,1534)(1526,1534)(1547,1534)(1567,1534)(1588,1534)(1608,1534)(1629,1534)(1649,1534)(1670,1534)(1691,1534)
\thinlines \path(1691,1534)(1711,1534)(1732,1534)(1752,1534)(1773,1534)(1793,1534)(1814,1534)(1835,1534)(1855,1534)(1876,1534)(1896,1534)(1917,1534)(1937,1534)(1958,1534)(1979,1534)(1999,1534)(2020,1534)(2040,1534)(2061,1534)(2081,1534)(2102,1534)(2123,1534)(2143,1534)(2164,1534)(2184,1534)(2205,1534)(2226,1534)(2246,1534)(2267,1534)(2287,1534)(2308,1534)(2328,1534)(2349,1534)(2370,1534)(2390,1534)(2411,1534)(2431,1534)(2452,1534)(2472,1534)(2493,1534)(2514,1534)(2534,1534)(2555,1534)(2575,1534)(2596,1534)(2616,1534)(2637,1534)(2658,1534)(2678,1534)(2699,1534)(2719,1534)
\color{green}
\thinlines \path(682,901)(682,901)(703,915)(724,928)(744,940)(765,952)(785,964)(806,975)(826,986)(847,996)(868,1006)(888,1015)(909,1024)(929,1033)(950,1042)(970,1050)(991,1058)(1012,1066)(1032,1073)(1053,1080)(1073,1087)(1094,1094)(1114,1101)(1135,1107)(1156,1113)(1176,1119)(1197,1125)(1217,1131)(1238,1137)(1258,1142)(1279,1147)(1300,1152)(1320,1157)(1341,1162)(1361,1167)(1382,1171)(1403,1176)(1423,1180)(1444,1185)(1464,1189)(1485,1193)(1505,1197)(1526,1201)(1547,1205)(1567,1208)(1588,1212)(1608,1216)(1629,1219)(1649,1222)(1670,1226)(1691,1229)
\thinlines \path(1691,1229)(1711,1232)(1732,1235)(1752,1238)(1773,1241)(1793,1244)(1814,1247)(1835,1250)(1855,1253)(1876,1255)(1896,1258)(1917,1261)(1937,1263)(1958,1266)(1979,1268)(1999,1271)(2020,1273)(2040,1275)(2061,1278)(2081,1280)(2102,1282)(2123,1284)(2143,1287)(2164,1289)(2184,1291)(2205,1293)(2226,1295)(2246,1297)(2267,1299)(2287,1301)(2308,1302)(2328,1304)(2349,1306)(2370,1308)(2390,1310)(2411,1311)(2431,1313)(2452,1315)(2472,1316)(2493,1318)(2514,1320)(2534,1321)(2555,1323)(2575,1324)(2596,1326)(2616,1327)(2637,1329)(2658,1330)(2678,1332)(2699,1333)(2719,1335)
\end{picture}
\caption{\label{fig6}The frequency gap~(\ref{eqgap}) for electromagnetic perturbations in $d=4$ for $r_H=100$ and $\ell = 1$: zeroth and first order analytical (eq.~(\ref{eq129})) compared with numerical data~\cite{CKL}.}
\end{center}
\end{figure}
For a large black hole, using~(\ref{eq107}), we obtain the spectrum
\be\label{eq129} \frac{\Delta\o_n}{r_H} \approx \frac{3\sqrt 3 (1-i\sqrt 3)}{4} \left( 1 - \frac{i}{4\pi n} + \dots \right) = 1.299 - 2.25i -\frac{0.179 + 0.103 i}{n} + \dots\ee
This analytical result is compared with numerical results~\cite{CKL} for $r_H=100$ in figure~\ref{fig6}.

\begin{figure}
\begin{center}
\setlength{\unitlength}{0.0750pt}
\begin{picture}(3000,1800)(0,0)
\footnotesize
\color{black}
\thicklines \path(328,249)(369,249)
\thicklines \path(2876,249)(2835,249)
\put(287,249){\makebox(0,0)[r]{ 1.65}}
\thicklines \path(328,459)(369,459)
\thicklines \path(2876,459)(2835,459)
\put(287,459){\makebox(0,0)[r]{ 1.7}}
\thicklines \path(328,669)(369,669)
\thicklines \path(2876,669)(2835,669)
\put(287,669){\makebox(0,0)[r]{ 1.75}}
\thicklines \path(328,879)(369,879)
\thicklines \path(2876,879)(2835,879)
\put(287,879){\makebox(0,0)[r]{ 1.8}}
\thicklines \path(328,1088)(369,1088)
\thicklines \path(2876,1088)(2835,1088)
\put(287,1088){\makebox(0,0)[r]{ 1.85}}
\thicklines \path(328,1298)(369,1298)
\thicklines \path(2876,1298)(2835,1298)
\put(287,1298){\makebox(0,0)[r]{ 1.9}}
\thicklines \path(328,1508)(369,1508)
\thicklines \path(2876,1508)(2835,1508)
\put(287,1508){\makebox(0,0)[r]{ 1.95}}
\thicklines \path(328,1718)(369,1718)
\thicklines \path(2876,1718)(2835,1718)
\put(287,1718){\makebox(0,0)[r]{ 2}}
\thicklines \path(328,249)(328,290)
\thicklines \path(328,1718)(328,1677)
\put(328,166){\makebox(0,0){ 0}}
\thicklines \path(583,249)(583,290)
\thicklines \path(583,1718)(583,1677)
\put(583,166){\makebox(0,0){ 2}}
\thicklines \path(838,249)(838,290)
\thicklines \path(838,1718)(838,1677)
\put(838,166){\makebox(0,0){ 4}}
\thicklines \path(1092,249)(1092,290)
\thicklines \path(1092,1718)(1092,1677)
\put(1092,166){\makebox(0,0){ 6}}
\thicklines \path(1347,249)(1347,290)
\thicklines \path(1347,1718)(1347,1677)
\put(1347,166){\makebox(0,0){ 8}}
\thicklines \path(1602,249)(1602,290)
\thicklines \path(1602,1718)(1602,1677)
\put(1602,166){\makebox(0,0){ 10}}
\thicklines \path(1857,249)(1857,290)
\thicklines \path(1857,1718)(1857,1677)
\put(1857,166){\makebox(0,0){ 12}}
\thicklines \path(2112,249)(2112,290)
\thicklines \path(2112,1718)(2112,1677)
\put(2112,166){\makebox(0,0){ 14}}
\thicklines \path(2366,249)(2366,290)
\thicklines \path(2366,1718)(2366,1677)
\put(2366,166){\makebox(0,0){ 16}}
\thicklines \path(2621,249)(2621,290)
\thicklines \path(2621,1718)(2621,1677)
\put(2621,166){\makebox(0,0){ 18}}
\thicklines \path(2876,249)(2876,290)
\thicklines \path(2876,1718)(2876,1677)
\put(2876,166){\makebox(0,0){ 20}}
\color{black}
\thicklines \path(328,249)(2876,249)(2876,1718)(328,1718)(328,249)
\put(1602,42){\makebox(0,0){$\Re\Delta\omega$}}
\color{red}
\put(455,378){\makebox(0,0){$\Diamond$}}
\put(583,1003){\makebox(0,0){$\Diamond$}}
\put(710,1215){\makebox(0,0){$\Diamond$}}
\put(838,1318){\makebox(0,0){$\Diamond$}}
\put(965,1377){\makebox(0,0){$\Diamond$}}
\put(1092,1416){\makebox(0,0){$\Diamond$}}
\put(1220,1443){\makebox(0,0){$\Diamond$}}
\put(1347,1463){\makebox(0,0){$\Diamond$}}
\put(1475,1478){\makebox(0,0){$\Diamond$}}
\put(1602,1490){\makebox(0,0){$\Diamond$}}
\put(1729,1500){\makebox(0,0){$\Diamond$}}
\put(1857,1508){\makebox(0,0){$\Diamond$}}
\put(1984,1514){\makebox(0,0){$\Diamond$}}
\put(2112,1520){\makebox(0,0){$\Diamond$}}
\put(2239,1525){\makebox(0,0){$\Diamond$}}
\put(2366,1529){\makebox(0,0){$\Diamond$}}
\put(2494,1533){\makebox(0,0){$\Diamond$}}
\put(2621,1536){\makebox(0,0){$\Diamond$}}
\put(2749,1539){\makebox(0,0){$\Diamond$}}
\color{blue}
\thinlines \path(455,1588)(455,1588)(479,1588)(502,1588)(525,1588)(548,1588)(571,1588)(594,1588)(618,1588)(641,1588)(664,1588)(687,1588)(710,1588)(733,1588)(757,1588)(780,1588)(803,1588)(826,1588)(849,1588)(872,1588)(896,1588)(919,1588)(942,1588)(965,1588)(988,1588)(1011,1588)(1034,1588)(1058,1588)(1081,1588)(1104,1588)(1127,1588)(1150,1588)(1173,1588)(1197,1588)(1220,1588)(1243,1588)(1266,1588)(1289,1588)(1312,1588)(1336,1588)(1359,1588)(1382,1588)(1405,1588)(1428,1588)(1451,1588)(1475,1588)(1498,1588)(1521,1588)(1544,1588)(1567,1588)(1590,1588)
\thinlines \path(1590,1588)(1614,1588)(1637,1588)(1660,1588)(1683,1588)(1706,1588)(1729,1588)(1753,1588)(1776,1588)(1799,1588)(1822,1588)(1845,1588)(1868,1588)(1892,1588)(1915,1588)(1938,1588)(1961,1588)(1984,1588)(2007,1588)(2031,1588)(2054,1588)(2077,1588)(2100,1588)(2123,1588)(2146,1588)(2170,1588)(2193,1588)(2216,1588)(2239,1588)(2262,1588)(2285,1588)(2308,1588)(2332,1588)(2355,1588)(2378,1588)(2401,1588)(2424,1588)(2447,1588)(2471,1588)(2494,1588)(2517,1588)(2540,1588)(2563,1588)(2586,1588)(2610,1588)(2633,1588)(2656,1588)(2679,1588)(2702,1588)(2725,1588)(2749,1588)
\color{green}
\thinlines \path(455,803)(455,803)(479,924)(502,1012)(525,1080)(548,1133)(571,1177)(594,1213)(618,1243)(641,1268)(664,1290)(687,1309)(710,1326)(733,1341)(757,1355)(780,1367)(803,1377)(826,1387)(849,1396)(872,1404)(896,1412)(919,1419)(942,1425)(965,1431)(988,1436)(1011,1442)(1034,1446)(1058,1451)(1081,1455)(1104,1459)(1127,1463)(1150,1466)(1173,1470)(1197,1473)(1220,1476)(1243,1479)(1266,1481)(1289,1484)(1312,1486)(1336,1489)(1359,1491)(1382,1493)(1405,1495)(1428,1497)(1451,1499)(1475,1501)(1498,1502)(1521,1504)(1544,1506)(1567,1507)(1590,1509)
\thinlines \path(1590,1509)(1614,1510)(1637,1511)(1660,1513)(1683,1514)(1706,1515)(1729,1517)(1753,1518)(1776,1519)(1799,1520)(1822,1521)(1845,1522)(1868,1523)(1892,1524)(1915,1525)(1938,1526)(1961,1527)(1984,1528)(2007,1528)(2031,1529)(2054,1530)(2077,1531)(2100,1531)(2123,1532)(2146,1533)(2170,1534)(2193,1534)(2216,1535)(2239,1536)(2262,1536)(2285,1537)(2308,1537)(2332,1538)(2355,1539)(2378,1539)(2401,1540)(2424,1540)(2447,1541)(2471,1541)(2494,1542)(2517,1542)(2540,1543)(2563,1543)(2586,1544)(2610,1544)(2633,1545)(2656,1545)(2679,1545)(2702,1546)(2725,1546)(2749,1547)
\end{picture}
\setlength{\unitlength}{0.0750pt}
\begin{picture}(3000,1800)(0,0)
\footnotesize
\color{black}
\thicklines \path(328,249)(369,249)
\thicklines \path(2876,249)(2835,249)
\put(287,249){\makebox(0,0)[r]{-2.52}}
\thicklines \path(328,412)(369,412)
\thicklines \path(2876,412)(2835,412)
\put(287,412){\makebox(0,0)[r]{-2.5}}
\thicklines \path(328,575)(369,575)
\thicklines \path(2876,575)(2835,575)
\put(287,575){\makebox(0,0)[r]{-2.48}}
\thicklines \path(328,739)(369,739)
\thicklines \path(2876,739)(2835,739)
\put(287,739){\makebox(0,0)[r]{-2.46}}
\thicklines \path(328,902)(369,902)
\thicklines \path(2876,902)(2835,902)
\put(287,902){\makebox(0,0)[r]{-2.44}}
\thicklines \path(328,1065)(369,1065)
\thicklines \path(2876,1065)(2835,1065)
\put(287,1065){\makebox(0,0)[r]{-2.42}}
\thicklines \path(328,1228)(369,1228)
\thicklines \path(2876,1228)(2835,1228)
\put(287,1228){\makebox(0,0)[r]{-2.4}}
\thicklines \path(328,1392)(369,1392)
\thicklines \path(2876,1392)(2835,1392)
\put(287,1392){\makebox(0,0)[r]{-2.38}}
\thicklines \path(328,1555)(369,1555)
\thicklines \path(2876,1555)(2835,1555)
\put(287,1555){\makebox(0,0)[r]{-2.36}}
\thicklines \path(328,1718)(369,1718)
\thicklines \path(2876,1718)(2835,1718)
\put(287,1718){\makebox(0,0)[r]{-2.34}}
\thicklines \path(328,249)(328,290)
\thicklines \path(328,1718)(328,1677)
\put(328,166){\makebox(0,0){ 0}}
\thicklines \path(583,249)(583,290)
\thicklines \path(583,1718)(583,1677)
\put(583,166){\makebox(0,0){ 2}}
\thicklines \path(838,249)(838,290)
\thicklines \path(838,1718)(838,1677)
\put(838,166){\makebox(0,0){ 4}}
\thicklines \path(1092,249)(1092,290)
\thicklines \path(1092,1718)(1092,1677)
\put(1092,166){\makebox(0,0){ 6}}
\thicklines \path(1347,249)(1347,290)
\thicklines \path(1347,1718)(1347,1677)
\put(1347,166){\makebox(0,0){ 8}}
\thicklines \path(1602,249)(1602,290)
\thicklines \path(1602,1718)(1602,1677)
\put(1602,166){\makebox(0,0){ 10}}
\thicklines \path(1857,249)(1857,290)
\thicklines \path(1857,1718)(1857,1677)
\put(1857,166){\makebox(0,0){ 12}}
\thicklines \path(2112,249)(2112,290)
\thicklines \path(2112,1718)(2112,1677)
\put(2112,166){\makebox(0,0){ 14}}
\thicklines \path(2366,249)(2366,290)
\thicklines \path(2366,1718)(2366,1677)
\put(2366,166){\makebox(0,0){ 16}}
\thicklines \path(2621,249)(2621,290)
\thicklines \path(2621,1718)(2621,1677)
\put(2621,166){\makebox(0,0){ 18}}
\thicklines \path(2876,249)(2876,290)
\thicklines \path(2876,1718)(2876,1677)
\put(2876,166){\makebox(0,0){ 20}}
\color{black}
\thicklines \path(328,249)(2876,249)(2876,1718)(328,1718)(328,249)
\put(1602,42){\makebox(0,0){$\Im\Delta\omega$}}
\color{red}
\put(455,797){\makebox(0,0){$\Diamond$}}
\put(583,1028){\makebox(0,0){$\Diamond$}}
\put(710,1195){\makebox(0,0){$\Diamond$}}
\put(838,1295){\makebox(0,0){$\Diamond$}}
\put(965,1359){\makebox(0,0){$\Diamond$}}
\put(1092,1404){\makebox(0,0){$\Diamond$}}
\put(1220,1436){\makebox(0,0){$\Diamond$}}
\put(1347,1461){\makebox(0,0){$\Diamond$}}
\put(1475,1480){\makebox(0,0){$\Diamond$}}
\put(1602,1496){\makebox(0,0){$\Diamond$}}
\put(1729,1509){\makebox(0,0){$\Diamond$}}
\put(1857,1519){\makebox(0,0){$\Diamond$}}
\put(1984,1528){\makebox(0,0){$\Diamond$}}
\put(2112,1536){\makebox(0,0){$\Diamond$}}
\put(2239,1543){\makebox(0,0){$\Diamond$}}
\put(2366,1549){\makebox(0,0){$\Diamond$}}
\put(2494,1554){\makebox(0,0){$\Diamond$}}
\put(2621,1558){\makebox(0,0){$\Diamond$}}
\put(2749,1563){\makebox(0,0){$\Diamond$}}
\color{blue}
\thinlines \path(455,1636)(455,1636)(479,1636)(502,1636)(525,1636)(548,1636)(571,1636)(594,1636)(618,1636)(641,1636)(664,1636)(687,1636)(710,1636)(733,1636)(757,1636)(780,1636)(803,1636)(826,1636)(849,1636)(872,1636)(896,1636)(919,1636)(942,1636)(965,1636)(988,1636)(1011,1636)(1034,1636)(1058,1636)(1081,1636)(1104,1636)(1127,1636)(1150,1636)(1173,1636)(1197,1636)(1220,1636)(1243,1636)(1266,1636)(1289,1636)(1312,1636)(1336,1636)(1359,1636)(1382,1636)(1405,1636)(1428,1636)(1451,1636)(1475,1636)(1498,1636)(1521,1636)(1544,1636)(1567,1636)(1590,1636)
\thinlines \path(1590,1636)(1614,1636)(1637,1636)(1660,1636)(1683,1636)(1706,1636)(1729,1636)(1753,1636)(1776,1636)(1799,1636)(1822,1636)(1845,1636)(1868,1636)(1892,1636)(1915,1636)(1938,1636)(1961,1636)(1984,1636)(2007,1636)(2031,1636)(2054,1636)(2077,1636)(2100,1636)(2123,1636)(2146,1636)(2170,1636)(2193,1636)(2216,1636)(2239,1636)(2262,1636)(2285,1636)(2308,1636)(2332,1636)(2355,1636)(2378,1636)(2401,1636)(2424,1636)(2447,1636)(2471,1636)(2494,1636)(2517,1636)(2540,1636)(2563,1636)(2586,1636)(2610,1636)(2633,1636)(2656,1636)(2679,1636)(2702,1636)(2725,1636)(2749,1636)
\color{green}
\thinlines \path(455,358)(455,358)(479,554)(502,699)(525,809)(548,896)(571,967)(594,1025)(618,1074)(641,1115)(664,1151)(687,1183)(710,1210)(733,1234)(757,1256)(780,1276)(803,1293)(826,1309)(849,1324)(872,1337)(896,1349)(919,1361)(942,1371)(965,1381)(988,1390)(1011,1398)(1034,1406)(1058,1413)(1081,1420)(1104,1426)(1127,1433)(1150,1438)(1173,1444)(1197,1449)(1220,1454)(1243,1458)(1266,1463)(1289,1467)(1312,1471)(1336,1475)(1359,1478)(1382,1482)(1405,1485)(1428,1488)(1451,1491)(1475,1494)(1498,1497)(1521,1500)(1544,1502)(1567,1505)(1590,1507)
\thinlines \path(1590,1507)(1614,1510)(1637,1512)(1660,1514)(1683,1516)(1706,1518)(1729,1520)(1753,1522)(1776,1524)(1799,1526)(1822,1527)(1845,1529)(1868,1531)(1892,1532)(1915,1534)(1938,1535)(1961,1537)(1984,1538)(2007,1539)(2031,1541)(2054,1542)(2077,1543)(2100,1544)(2123,1546)(2146,1547)(2170,1548)(2193,1549)(2216,1550)(2239,1551)(2262,1552)(2285,1553)(2308,1554)(2332,1555)(2355,1556)(2378,1557)(2401,1558)(2424,1559)(2447,1560)(2471,1560)(2494,1561)(2517,1562)(2540,1563)(2563,1564)(2586,1564)(2610,1565)(2633,1566)(2656,1566)(2679,1567)(2702,1568)(2725,1568)(2749,1569)
\end{picture}
\caption{\label{fig7}The frequency gap~(\ref{eqgap}) for electromagnetic perturbations in $d=4$ for $r_H=1$ and $\ell = 1$: zeroth and first order analytical (eq.~(\ref{eq130})) compared with numerical data~\cite{CKL}.}
\end{center}
\end{figure}
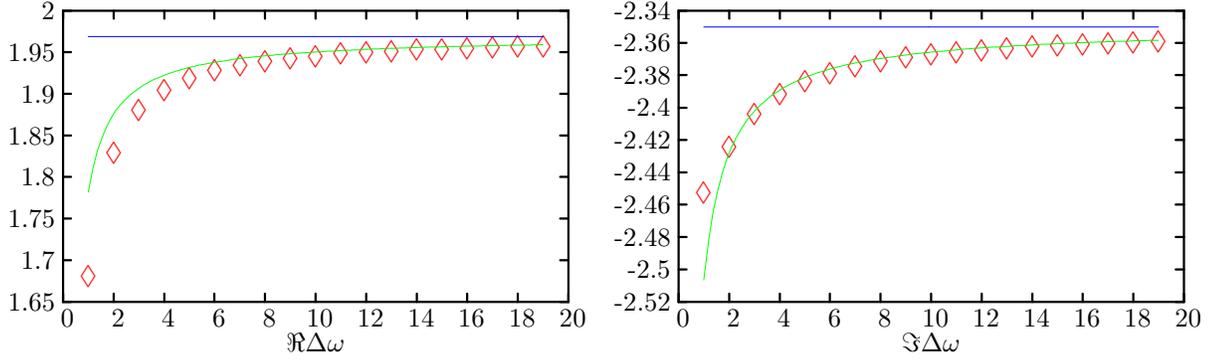
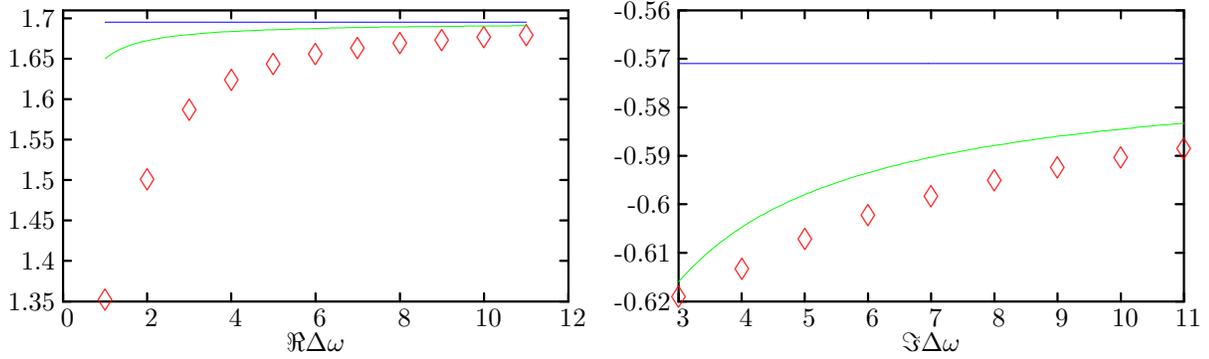
\begin{figure}
\begin{center}
\setlength{\unitlength}{0.0750pt}
\begin{picture}(3000,1800)(0,0)
\footnotesize
\color{black}
\thicklines \path(328,249)(369,249)
\thicklines \path(2876,249)(2835,249)
\put(287,249){\makebox(0,0)[r]{ 1.35}}
\thicklines \path(328,453)(369,453)
\thicklines \path(2876,453)(2835,453)
\put(287,453){\makebox(0,0)[r]{ 1.4}}
\thicklines \path(328,657)(369,657)
\thicklines \path(2876,657)(2835,657)
\put(287,657){\makebox(0,0)[r]{ 1.45}}
\thicklines \path(328,861)(369,861)
\thicklines \path(2876,861)(2835,861)
\put(287,861){\makebox(0,0)[r]{ 1.5}}
\thicklines \path(328,1065)(369,1065)
\thicklines \path(2876,1065)(2835,1065)
\put(287,1065){\makebox(0,0)[r]{ 1.55}}
\thicklines \path(328,1269)(369,1269)
\thicklines \path(2876,1269)(2835,1269)
\put(287,1269){\makebox(0,0)[r]{ 1.6}}
\thicklines \path(328,1473)(369,1473)
\thicklines \path(2876,1473)(2835,1473)
\put(287,1473){\makebox(0,0)[r]{ 1.65}}
\thicklines \path(328,1677)(369,1677)
\thicklines \path(2876,1677)(2835,1677)
\put(287,1677){\makebox(0,0)[r]{ 1.7}}
\thicklines \path(328,249)(328,290)
\thicklines \path(328,1718)(328,1677)
\put(328,166){\makebox(0,0){ 0}}
\thicklines \path(753,249)(753,290)
\thicklines \path(753,1718)(753,1677)
\put(753,166){\makebox(0,0){ 2}}
\thicklines \path(1177,249)(1177,290)
\thicklines \path(1177,1718)(1177,1677)
\put(1177,166){\makebox(0,0){ 4}}
\thicklines \path(1602,249)(1602,290)
\thicklines \path(1602,1718)(1602,1677)
\put(1602,166){\makebox(0,0){ 6}}
\thicklines \path(2027,249)(2027,290)
\thicklines \path(2027,1718)(2027,1677)
\put(2027,166){\makebox(0,0){ 8}}
\thicklines \path(2451,249)(2451,290)
\thicklines \path(2451,1718)(2451,1677)
\put(2451,166){\makebox(0,0){ 10}}
\thicklines \path(2876,249)(2876,290)
\thicklines \path(2876,1718)(2876,1677)
\put(2876,166){\makebox(0,0){ 12}}
\color{black}
\thicklines \path(328,249)(2876,249)(2876,1718)(328,1718)(328,249)
\put(1602,42){\makebox(0,0){$\Re\Delta\omega$}}
\color{red}
\put(540,258){\makebox(0,0){$\Diamond$}}
\put(753,866){\makebox(0,0){$\Diamond$}}
\put(965,1214){\makebox(0,0){$\Diamond$}}
\put(1177,1364){\makebox(0,0){$\Diamond$}}
\put(1390,1446){\makebox(0,0){$\Diamond$}}
\put(1602,1496){\makebox(0,0){$\Diamond$}}
\put(1814,1529){\makebox(0,0){$\Diamond$}}
\put(2027,1552){\makebox(0,0){$\Diamond$}}
\put(2239,1569){\makebox(0,0){$\Diamond$}}
\put(2451,1582){\makebox(0,0){$\Diamond$}}
\put(2664,1591){\makebox(0,0){$\Diamond$}}
\color{blue}
\thinlines \path(540,1657)(540,1657)(562,1657)(583,1657)(605,1657)(626,1657)(648,1657)(669,1657)(690,1657)(712,1657)(733,1657)(755,1657)(776,1657)(798,1657)(819,1657)(841,1657)(862,1657)(883,1657)(905,1657)(926,1657)(948,1657)(969,1657)(991,1657)(1012,1657)(1034,1657)(1055,1657)(1077,1657)(1098,1657)(1119,1657)(1141,1657)(1162,1657)(1184,1657)(1205,1657)(1227,1657)(1248,1657)(1270,1657)(1291,1657)(1312,1657)(1334,1657)(1355,1657)(1377,1657)(1398,1657)(1420,1657)(1441,1657)(1463,1657)(1484,1657)(1505,1657)(1527,1657)(1548,1657)(1570,1657)(1591,1657)
\thinlines \path(1591,1657)(1613,1657)(1634,1657)(1656,1657)(1677,1657)(1699,1657)(1720,1657)(1741,1657)(1763,1657)(1784,1657)(1806,1657)(1827,1657)(1849,1657)(1870,1657)(1892,1657)(1913,1657)(1934,1657)(1956,1657)(1977,1657)(1999,1657)(2020,1657)(2042,1657)(2063,1657)(2085,1657)(2106,1657)(2127,1657)(2149,1657)(2170,1657)(2192,1657)(2213,1657)(2235,1657)(2256,1657)(2278,1657)(2299,1657)(2321,1657)(2342,1657)(2363,1657)(2385,1657)(2406,1657)(2428,1657)(2449,1657)(2471,1657)(2492,1657)(2514,1657)(2535,1657)(2556,1657)(2578,1657)(2599,1657)(2621,1657)(2642,1657)(2664,1657)
\color{green}
\thinlines \path(540,1473)(540,1473)(562,1490)(583,1504)(605,1516)(626,1526)(648,1535)(669,1542)(690,1549)(712,1555)(733,1561)(755,1565)(776,1570)(798,1574)(819,1577)(841,1581)(862,1584)(883,1587)(905,1589)(926,1592)(948,1594)(969,1596)(991,1598)(1012,1600)(1034,1602)(1055,1603)(1077,1605)(1098,1606)(1119,1608)(1141,1609)(1162,1610)(1184,1611)(1205,1612)(1227,1613)(1248,1614)(1270,1615)(1291,1616)(1312,1617)(1334,1618)(1355,1619)(1377,1620)(1398,1620)(1420,1621)(1441,1622)(1463,1622)(1484,1623)(1505,1624)(1527,1624)(1548,1625)(1570,1625)(1591,1626)
\thinlines \path(1591,1626)(1613,1626)(1634,1627)(1656,1627)(1677,1628)(1699,1628)(1720,1629)(1741,1629)(1763,1630)(1784,1630)(1806,1630)(1827,1631)(1849,1631)(1870,1632)(1892,1632)(1913,1632)(1934,1633)(1956,1633)(1977,1633)(1999,1633)(2020,1634)(2042,1634)(2063,1634)(2085,1635)(2106,1635)(2127,1635)(2149,1635)(2170,1636)(2192,1636)(2213,1636)(2235,1636)(2256,1637)(2278,1637)(2299,1637)(2321,1637)(2342,1637)(2363,1638)(2385,1638)(2406,1638)(2428,1638)(2449,1638)(2471,1639)(2492,1639)(2514,1639)(2535,1639)(2556,1639)(2578,1639)(2599,1640)(2621,1640)(2642,1640)(2664,1640)
\end{picture}
\setlength{\unitlength}{0.0750pt}
\begin{picture}(3000,1800)(0,0)
\footnotesize
\color{black}
\thicklines \path(328,249)(369,249)
\thicklines \path(2876,249)(2835,249)
\put(287,249){\makebox(0,0)[r]{-0.62}}
\thicklines \path(328,494)(369,494)
\thicklines \path(2876,494)(2835,494)
\put(287,494){\makebox(0,0)[r]{-0.61}}
\thicklines \path(328,739)(369,739)
\thicklines \path(2876,739)(2835,739)
\put(287,739){\makebox(0,0)[r]{-0.6}}
\thicklines \path(328,984)(369,984)
\thicklines \path(2876,984)(2835,984)
\put(287,984){\makebox(0,0)[r]{-0.59}}
\thicklines \path(328,1228)(369,1228)
\thicklines \path(2876,1228)(2835,1228)
\put(287,1228){\makebox(0,0)[r]{-0.58}}
\thicklines \path(328,1473)(369,1473)
\thicklines \path(2876,1473)(2835,1473)
\put(287,1473){\makebox(0,0)[r]{-0.57}}
\thicklines \path(328,1718)(369,1718)
\thicklines \path(2876,1718)(2835,1718)
\put(287,1718){\makebox(0,0)[r]{-0.56}}
\thicklines \path(328,249)(328,290)
\thicklines \path(328,1718)(328,1677)
\put(328,166){\makebox(0,0){ 3}}
\thicklines \path(647,249)(647,290)
\thicklines \path(647,1718)(647,1677)
\put(647,166){\makebox(0,0){ 4}}
\thicklines \path(965,249)(965,290)
\thicklines \path(965,1718)(965,1677)
\put(965,166){\makebox(0,0){ 5}}
\thicklines \path(1284,249)(1284,290)
\thicklines \path(1284,1718)(1284,1677)
\put(1284,166){\makebox(0,0){ 6}}
\thicklines \path(1602,249)(1602,290)
\thicklines \path(1602,1718)(1602,1677)
\put(1602,166){\makebox(0,0){ 7}}
\thicklines \path(1921,249)(1921,290)
\thicklines \path(1921,1718)(1921,1677)
\put(1921,166){\makebox(0,0){ 8}}
\thicklines \path(2239,249)(2239,290)
\thicklines \path(2239,1718)(2239,1677)
\put(2239,166){\makebox(0,0){ 9}}
\thicklines \path(2558,249)(2558,290)
\thicklines \path(2558,1718)(2558,1677)
\put(2558,166){\makebox(0,0){ 10}}
\thicklines \path(2876,249)(2876,290)
\thicklines \path(2876,1718)(2876,1677)
\put(2876,166){\makebox(0,0){ 11}}
\color{black}
\thicklines \path(328,249)(2876,249)(2876,1718)(328,1718)(328,249)
\put(1602,42){\makebox(0,0){$\Im\Delta\omega$}}
\color{red}
\put(328,272){\makebox(0,0){$\Diamond$}}
\put(647,413){\makebox(0,0){$\Diamond$}}
\put(965,560){\makebox(0,0){$\Diamond$}}
\put(1284,683){\makebox(0,0){$\Diamond$}}
\put(1602,781){\makebox(0,0){$\Diamond$}}
\put(1921,860){\makebox(0,0){$\Diamond$}}
\put(2239,923){\makebox(0,0){$\Diamond$}}
\put(2558,975){\makebox(0,0){$\Diamond$}}
\put(2876,1019){\makebox(0,0){$\Diamond$}}
\color{blue}
\thinlines \path(328,1449)(328,1449)(354,1449)(379,1449)(405,1449)(431,1449)(457,1449)(482,1449)(508,1449)(534,1449)(560,1449)(585,1449)(611,1449)(637,1449)(663,1449)(688,1449)(714,1449)(740,1449)(766,1449)(791,1449)(817,1449)(843,1449)(868,1449)(894,1449)(920,1449)(946,1449)(971,1449)(997,1449)(1023,1449)(1049,1449)(1074,1449)(1100,1449)(1126,1449)(1152,1449)(1177,1449)(1203,1449)(1229,1449)(1255,1449)(1280,1449)(1306,1449)(1332,1449)(1357,1449)(1383,1449)(1409,1449)(1435,1449)(1460,1449)(1486,1449)(1512,1449)(1538,1449)(1563,1449)(1589,1449)
\thinlines \path(1589,1449)(1615,1449)(1641,1449)(1666,1449)(1692,1449)(1718,1449)(1744,1449)(1769,1449)(1795,1449)(1821,1449)(1847,1449)(1872,1449)(1898,1449)(1924,1449)(1949,1449)(1975,1449)(2001,1449)(2027,1449)(2052,1449)(2078,1449)(2104,1449)(2130,1449)(2155,1449)(2181,1449)(2207,1449)(2233,1449)(2258,1449)(2284,1449)(2310,1449)(2336,1449)(2361,1449)(2387,1449)(2413,1449)(2438,1449)(2464,1449)(2490,1449)(2516,1449)(2541,1449)(2567,1449)(2593,1449)(2619,1449)(2644,1449)(2670,1449)(2696,1449)(2722,1449)(2747,1449)(2773,1449)(2799,1449)(2825,1449)(2850,1449)(2876,1449)
\color{green}
\thinlines \path(328,347)(328,347)(354,376)(379,403)(405,429)(431,454)(457,478)(482,500)(508,522)(534,542)(560,562)(585,581)(611,599)(637,616)(663,633)(688,649)(714,664)(740,679)(766,693)(791,707)(817,720)(843,733)(868,745)(894,757)(920,768)(946,780)(971,790)(997,801)(1023,811)(1049,821)(1074,830)(1100,839)(1126,848)(1152,857)(1177,865)(1203,874)(1229,882)(1255,889)(1280,897)(1306,904)(1332,911)(1357,918)(1383,925)(1409,932)(1435,938)(1460,944)(1486,951)(1512,957)(1538,962)(1563,968)(1589,974)
\thinlines \path(1589,974)(1615,979)(1641,985)(1666,990)(1692,995)(1718,1000)(1744,1005)(1769,1009)(1795,1014)(1821,1019)(1847,1023)(1872,1028)(1898,1032)(1924,1036)(1949,1040)(1975,1044)(2001,1048)(2027,1052)(2052,1056)(2078,1060)(2104,1063)(2130,1067)(2155,1070)(2181,1074)(2207,1077)(2233,1081)(2258,1084)(2284,1087)(2310,1090)(2336,1093)(2361,1096)(2387,1099)(2413,1102)(2438,1105)(2464,1108)(2490,1111)(2516,1114)(2541,1116)(2567,1119)(2593,1122)(2619,1124)(2644,1127)(2670,1129)(2696,1132)(2722,1134)(2747,1137)(2773,1139)(2799,1141)(2825,1144)(2850,1146)(2876,1148)
\end{picture}
\caption{\label{fig8}The frequency gap~(\ref{eqgap}) for electromagnetic perturbations in $d=4$ for $r_H=0.2$ and $\ell = 1$: zeroth and first order analytical (eq.~(\ref{eq131})) compared with numerical data~\cite{CKL}.}
\end{center}
\end{figure}

Using eqs.~(\ref{eqmur}), (\ref{eqrm}), (\ref{eqrs}) and (\ref{eqEMo1}), we obtain the spectrum of an intermediate black hole, $r_H =1$,
(see figure~\ref{fig7})
\be\label{eq130} \o_n = (1.969-2.350i)n - (0.187+0.1567i)\ln n + \dots \ee
and for a small black hole, $r_H=0.2$, (see figure~\ref{fig8})
\be\label{eq131} \o_n = (1.695-0.571i) n - (0.045+0.135i)\ln n + \dots \ee
All first-order analytical results are in good agreement with numerical results~\cite{CKL}.

\section{Conclusions}\label{sect4}

We studied quasi-normal modes for Schwarzschild black holes in asymptotically
AdS spaces of arbitrary dimension.
We obtained analytical expressions by solving the wave equation perturbatively,
including first-order corrections.
We studied scalar, electromagnetic and gravitational perturbations.
In the case of massive scalar perturbations, we extended the method proposed
in~\cite{Siopsis} for large black holes and derived explicit expressions of
quasi-normal frequencies for black holes of arbitrary size as a perturbative
expansion in $1/m$, where $m$ is the mass of the perturbation.

This method is not directly applicable to massless modes, because the
perturbative expansion fails as $m\to 0$.
Instead, we obtained perturbative expansions of gravitational and electromagnetic modes
by extending the method proposed in~\cite{SuSi} for asymptotically flat spaces.
The perturbative expansion was based on zeroth-order results obtained in~\cite{CNS,NS}.
We showed that our analytical results were in good agreement with numerical data~\cite{CKL}.
In the case of electromagnetic perturbations, zeroth-order expressions do not
yield finite quasi-normal frequencies, because the effective potential vanishes~\cite{NS}.
By including first-order effects, we were able to arrive at finite analytical
expressions with logarithmic sub-leading contributions.

It would be interesting to extend these results to other types of black holes
and also understand their implications on the AdS/CFT correspondence.

\section*{Acknowledgments}
This work was supported in part by the US Department of Energy under grant DE-FG05-91ER40627.
The work of S.~M.~was partially supported by the Thailand Research
Fund.
S.~M.~also gratefully acknowledges the hospitality of the Department of Physics and
Astronomy at the University of Tennessee where part of the work was performed.


\end{document}